\documentclass[sigconf]{acmart}
\usepackage{booktabs,tabularx,enumitem,ragged2e,makecell}
\usepackage{listings}
\usepackage{xcolor}
\newcolumntype{L}{>{\RaggedRight\arraybackslash}X}
\usepackage{tabularx}
\usepackage{tikz}
\usepackage{soul}
\usepackage{url}
\usepackage[utf8]{inputenc}
\usepackage{graphicx}
\usepackage{amsmath}
\usepackage{amsthm}
\usepackage{booktabs}
\usepackage{algorithm}
\usepackage{algorithmic}
\usepackage{float}
\usepackage{array}
\usepackage{csquotes}
\usepackage{placeins}
\usepackage{longtable}

\usepackage{ltablex}

\keepXColumns

\graphicspath{{figures/}}
\usepackage{subfig}
\usepackage{multirow}
\usepackage{bm}
\usepackage{color}
\definecolor{Gray}{gray}{0.935}

\setcopyright{none}
\renewcommand\footnotetextcopyrightpermission[1]{}
\begin{document}
\pagestyle{plain}


\newcommand{\sahand}[1]{\textcolor{red}{#1-sahand}}
\newcommand{\inna}[1]{\textcolor{blue}{#1-Inna}}
\newcommand{\tim}[1]{\textcolor{blue}{#1-Tim}}

\newcommand{\xhdr}[1]{\vspace{1.6mm}\noindent{{\bf #1.}}}

\definecolor{valbest}{HTML}{d9ead3}
\newcommand{\valbest}[1]{\colorbox{valbest}{#1}}
\definecolor{valgood}{HTML}{d0e0e3}
\newcommand{\valgood}[1]{\colorbox{valgood}{#1}}
\definecolor{valmid}{HTML}{fce5cd}
\newcommand{\valmid}[1]{\colorbox{valmid}{#1}}
\definecolor{valbad}{HTML}{ead1dc}
\newcommand{\valbad}[1]{\colorbox{valbad}{#1}}

\newcommand{\valingood}[1]{\begingroup\setlength{\fboxsep}{2pt}
\colorbox{valgood}{#1}
\endgroup
}

\newcommand{\valinbad}[1]{\begingroup\setlength{\fboxsep}{2pt}
\colorbox{valbad}{#1}
\endgroup
}

\newcommand{\Hone}{$\mathbf{H_1}$}
\newcommand{\Htwo}{$\mathbf{H_2}$}
\newcommand{\Hthree}{$\mathbf{H_3}$}
\newcommand{\Hfour}{$\mathbf{H_4}$}
\newcommand{\Hfive}{$\mathbf{H_5}$}

\newcolumntype{M}[1]{>{\centering\arraybackslash}m{#1}}


\title[CandorMD]{CandorMD: An AI-Assisted Audio Simulation and Feedback System for Training Clinicians for Medical Error Disclosure}


\author{Inna Wanyin Lin}
\affiliation{%
  \institution{University of Washington}
  \city{Seattle}
\state{WA}
  \country{USA}
}

\author{Sahand Sabour}
\authornote{Work done during tenure as a visiting researcher at the University of Washington.}
\affiliation{%
  \institution{Tsinghua University}
    \city{Beijing}
  \country{China}
}

\author{Hong Sng}
\affiliation{%
  \institution{University of Washington}
    \city{Seattle}
\state{WA}
  \country{USA}
}

\author{Maxine Chan}
\affiliation{%
  \institution{University of Washington}
    \city{Seattle}
\state{WA}
  \country{USA}
}

\author{Minlie Huang}
\affiliation{%
  \institution{Tsinghua University}
  \city{Beijing}
  \country{China}}
  
\author{Andrew White}
\affiliation{%
  \institution{University of Washington}
    \city{Seattle}
\state{WA}
  \country{USA}
}
\author{Tim Althoff}
\affiliation{%
  \institution{University of Washington}
    \city{Seattle}
\state{WA}
  \country{USA}
}

\renewcommand{\shortauthors}{Lin et al.}

\settopmatter{authorsperrow=4}




\begin{abstract}Clinicians are expected to disclose harmful medical errors to patients and families in line with ethical, regulatory, and patient care standards, yet these conversations remain challenging because of their emotional complexity and limited training opportunities. Most physicians still learn primarily through lectures and observation, while static video tools--though available--are underused, lack adaptability across specialties, and deliver delayed, generic feedback. These gaps restrict skill development, reduce self-efficacy, and contribute to avoidance of disclosure conversations, ultimately compromising patient care and eroding trust. To address these needs, we designed CandorMD—an AI-assisted simulation system that provides real-time practice, actionable feedback, and diverse practice environments tailored to individual learning needs. We conducted semi-structured interviews with physicians, risk managers, patient advocates, and communication experts to understand current practices, identify gaps, and collect feedback on CandorMD. Based on these insights, we present findings and design recommendations for the future of AI-supported medical communication training.
\end{abstract}

\maketitle

\section{Introduction}
\label{introduction}

\begin{figure*}[ht]
    \centering
    \includegraphics[width=\textwidth]{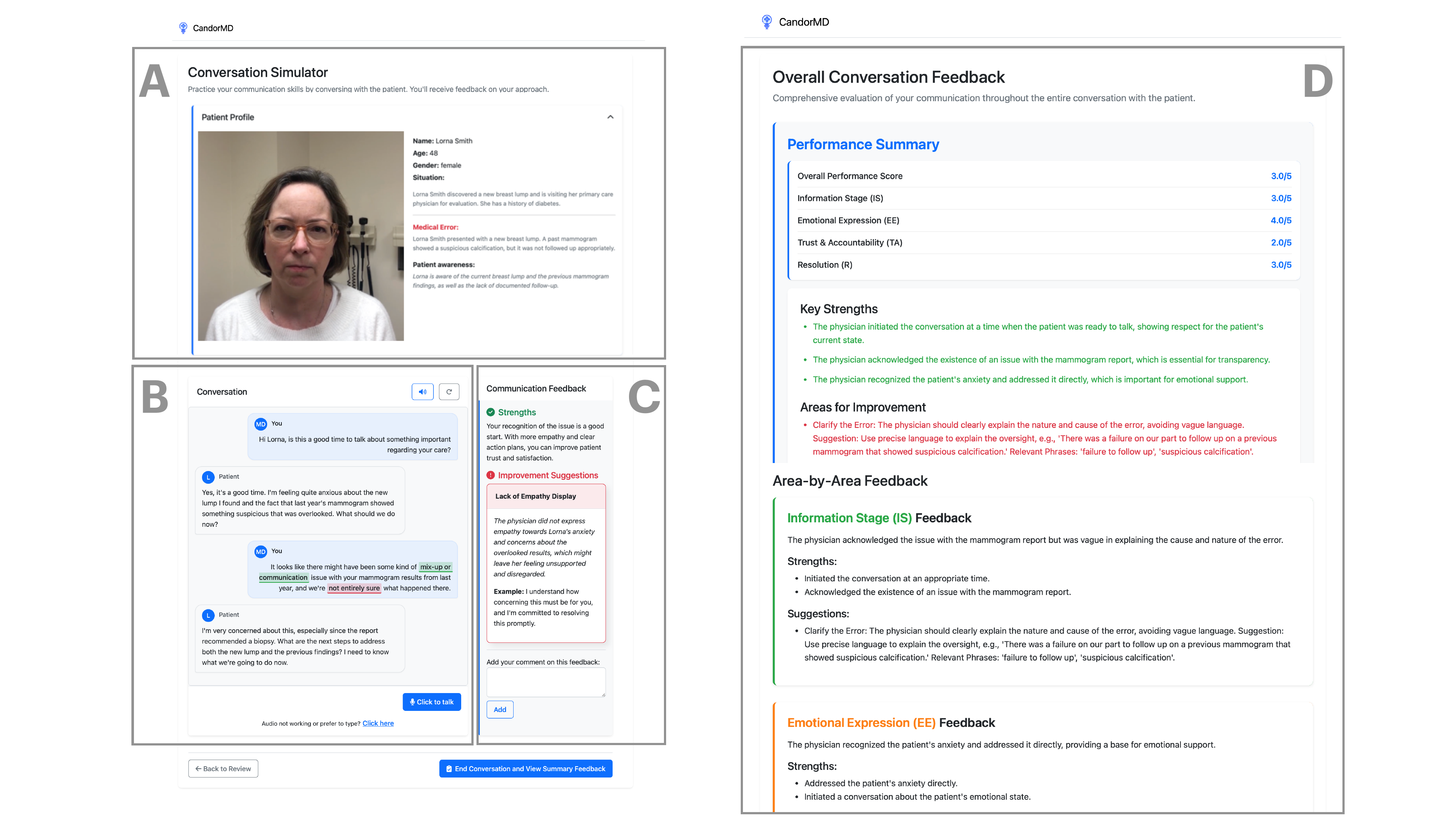}
    \caption{CandorMD core components. CandorMD is an audio-based simulation and feedback prototype for practicing medical error disclosure. Users begin by selecting or describing a case that includes a patient profile, error scenario, and patient knowledge state (A). They then engage in a simulated conversation (B), during which the system provides turn-by-turn feedback highlighting strengths and areas for improvement (C). Afterward, users receive an overall feedback report with overall performance, strengths, improvement suggestions, and area-specific feedback (D). Example images use actors portraying patients, included with consent. See Section~\ref{sec:demo_design} for further details.}
    \label{fig:overview}
\end{figure*}

Harmful medical errors are estimated to affect millions of patients globally~\cite{AhsaniEstahbanati2022}. When they occur, clinicians face an ethical, regulatory, and professional obligation to inform affected patients and their families about the error~\cite{Gallagher2007}. Yet, one of healthcare's most critical conversations—disclosing these errors to patients and families—remains among the most challenging tasks for clinicians to navigate effectively~\cite{gallagher2003patients, Bell2010, White2008}. These conversations serve multiple essential functions~\cite{kaldjian2021communication, surbone2012onclogists, mazor2004communicating, Mazor2004}: they provide patients with the necessary information for informed decision-making, provide emotional support during a period of distress, attempt to restore trust in the physician-patient relationship, and demonstrate accountability that forms the foundation of ethical medical practice.

The importance of effective error disclosure extends far beyond individual patient encounters. Research has demonstrated that transparent, empathetic communication about errors, paired with institutional efforts to provide reconciliation, can significantly improve patient satisfaction, strengthen rather than damage the therapeutic relationship between clinicians and patients, and reduce the likelihood of malpractice litigation~\cite{wu1999handling, Kachalia2018, Kachalia2010}. Furthermore, healthcare institutions increasingly recognize that fostering a culture of transparency around medical errors is essential for organizational learning, quality improvement, and maintaining public trust~\cite{Gallagher2020}. As healthcare systems worldwide implement disclosure policies and regulatory bodies establish disclosure requirements~\cite{cms2024medicare}, the ability to conduct these conversations competently has become not just ethically important but professionally essential.

Despite this recognized importance, error disclosure conversations remain exceptionally difficult for clinicians to master and execute effectively. The emotional complexity of these interactions presents significant challenges~\cite{kaldjian2007disclosing, surbone2012onclogists}: clinicians must simultaneously manage their own feelings of guilt, shame, and professional vulnerability while providing compassionate support to patients and families who may be experiencing anger, disappointment, or grief~\cite{WhiteCoping2011}. The conversations require a delicate balance of expressing accountability without speculation, and providing clear information about what happened while managing uncertainty about long-term consequences~\cite{kaldjian2021communication, miziara2025recognition}. These multifaceted demands make error disclosure fundamentally different from routine medical consultations, requiring specialized communication skills that many clinicians feel unprepared to deploy.

Current approaches to training clinicians in error disclosure conversations suffer from several critical limitations that have prevented the widespread development of competency in this essential skill. Most medical schools and residency programs rely heavily on didactic lectures and passive observation of experienced colleagues, methods that provide limited opportunities for active practice and skill development~\cite{2021CLERreport}. While a few institutions have introduced standardized patient encounters or role-playing exercises, these approaches are resource-intensive, difficult to scale, and often provide only generic scenarios that may not reflect the specific contexts and specialties where individual clinicians will practice~\cite{McDonough2017, Stroud2013}. At many hospitals, risk managers offer to provide just-in-time coaching to clinicians before some disclosure conversations~\cite{White2017}, but this labor-intensive approach is only used by a fraction of clinicians following very serious cases. Recent technological advances have led to the development of static video prompts that allow residents to practice formulating responses to disclosure scenarios~\cite{bauchat2016communication, grossniklaus2025error, White2022prepost, White2024}. However, these resources remain significantly underutilized and suffer from fundamental design limitations, including the lack of adaptability to diverse medical specialties, the inability to respond dynamically to learner inputs, and reliance on delayed, generic feedback that fails to address individual learning needs or specific response strategies\cite{Grossniklaus2025Faculty}.


Recent work demonstrates that AI, and specifically language models, can support individuals in learning and practicing cognitively demanding skills by offering interactive, self-guided interventions. For example, language models can provide suggestions to help users cognitively reframe negative thoughts~\cite{sharma-etal-2023-cognitive} and improve interpersonal skills~\cite{lin2024imbue,shaikh2024rehearsal}. These capabilities highlight how AI can extend access to practice opportunities that are otherwise limited in traditional training contexts. At the same time, these studies underscore the importance of developing such applications in close collaboration with domain expertise, particularly in high-stakes contexts such as healthcare.

Based on previous work to develop the video-based communication tool and decades of experience preparing clinical professionals for disclosure conversations, our research team developed CandorMD, an AI-assisted simulation and feedback system specifically designed to provide flexible, just-in-time support for learning disclosure skills through audio communication (Figure~\ref{fig:overview}).  Unlike existing approaches, CandorMD creates diverse, scenario-specific practice environments that can adapt to individual learning needs, specialty contexts, and varying personas of the simulated patient or family members. The system provides immediate, contextualized feedback and allows learners to practice repeatedly with scenarios that evolve based on their responses, more closely mimicking the dynamic nature of actual disclosure conversations.

To gain a more holistic understanding of the current practice and gaps in medical error disclosure conversations, we conducted semi-structured interviews with design probe with 12 key informants representing an ecosystem of medical error disclosure training and research. Our interview participants included practicing clinicians from multiple specialties, hospital risk managers responsible for developing and implementing disclosure policies and procedures, patient advocates who work with families affected by medical errors, and communication experts specializing in healthcare interactions. Through this study, we identified specific opportunities for technological innovation to better support skill development in error disclosure conversations.

After identifying the current landscape and opportunities, we examined how our prototype and potential adaptations could address these gaps. We formatively evaluated CandorMD with participants, gathering feedback on the system's technical implementation and its potential for real-world use. These insights informed design recommendations for future AI-supported tools for medical communication training.

Overall, we explored four research questions (RQ) in this study.
\begin{itemize}
    \item \textbf{RQ1 -- Key Factors}: What interpersonal, organizational, and regulatory factors shape whether and how clinicians disclose medical errors?
    \item \textbf{RQ2 -- Current Practices}: How do clinicians currently prepare for disclosure conversations, and what are the current approaches that facilitate long-term learning?
    \item \textbf{RQ3 -- Challenges and Shortcomings}: What are the challenges associated with disclosing medical errors, and what are the shortcomings of current methods for learning and preparation?
    \item \textbf{RQ4 -- Technology-Supported Training}: How can technology support preparation and learning for disclosure, as explored through a prototype design probe?
\end{itemize}

Our interviews underscored that effective disclosure relies on trust, empathy, and transparency. These disclosure conversations are not a single event but rather an ongoing process requiring both immediate and continuous communication. Clinicians described the dual challenge of regulating their own emotions while responding to patients' strong emotions. Existing training and feedback mechanisms were widely viewed as inconsistent and insufficient, pointing to opportunities for systems that enable realistic practice, deliver timely and actionable feedback, and scaffold support across the longitudinal course of disclosure conversations.

 While our current system focuses on error disclosure conversations, the design principles and technical approaches extend to other challenging healthcare communication scenarios. Our work contributes to research on AI-supported communication training, multimodal patient and persona simulation, and demonstrates how thoughtful integration of AI can complement traditional methods and address persistent gaps in medical education. Specifically, we make three contributions: (1) detailed insights into training needs and challenges in medical error disclosure, informed by interviews with diverse stakeholders; (2) the design and implementation of CandorMD as a design probe addressing limitations of current training approaches; and (3) empirically grounded design recommendations for developing AI-supported communication training tools that extend beyond medicine.

\section{Related Work}
\label{sec:related-work}

Our work builds upon previous research on medical error disclosure training and current practice (Section~\ref{subsec:related-disclosure-training}), human-AI collaboration for healthcare communication training (Section~\ref{subsec:related-ai-healthcare}), and the design of human-language model interaction systems (Section~\ref{subsec:related-human-language-interaction}).

\subsection{Medical Error Disclosure Training and Current Practice}
\label{subsec:related-disclosure-training}

Medical error disclosure represents one of the most challenging communication scenarios that physicians encounter, yet training approaches have remained largely unchanged for decades. Traditional medical education relies heavily on didactic instruction and observational learning, with limited opportunities for active practice in disclosure conversations~\cite{gallagher2006choosing,mazor2004health}. While some institutions have used standardized patients for disclosure training, this approach is resource-intensive and difficult to scale across diverse medical specialties and error contexts~\cite{wu2013disclosure}.

Recent advances in disclosure training have centered on Video-based Communication Assessment (VCA) tools, which represent the current state-of-the-art for structured disclosure practice. White et al.~\cite{white2022video, white2024crowdsourced} demonstrated that VCA approaches enable residents to practice responses to pre-recorded disclosure scenarios and receive feedback from laypeople as patient surrogates. Their work established that achieving adequate assessment reliability requires at least 12 raters and 9 vignettes per evaluation, highlighting the substantial resource requirements inherent in current training methods.

Despite these advances, existing VCA implementations suffer from fundamental limitations that constrain their effectiveness and adoption. The static nature of video scenarios prevents adaptive conversation flow, meaning residents cannot experience the dynamic reciprocal dialogue that characterizes actual disclosure conversations. Patient responses in real disclosure encounters evolve based on physician communication choices, emotional attunement, and disclosure technique—dynamics that cannot be captured through pre-recorded content~\cite{Triola2006, Yoon2016}. 

Furthermore, current approaches rely on delayed feedback from human evaluators, creating temporal gaps between practice and guidance that reduce learning effectiveness. The coordination requirements for multiple raters limit opportunities for just-in-time training when residents encounter actual disclosure situations. Additionally, existing VCA implementations offer limited specialty-specific customization, providing generic scenarios that may not reflect the diverse error types, patient populations, and institutional contexts encountered across medical domains~\cite{Grichanik2017, JunodPerron2016, Yang2023, Kononowicz2019}.

Assessment approaches in current disclosure training also face significant constraints. Crowdsourced rating methods, while more scalable than expert evaluation, introduce variability in feedback quality and lack the contextual medical knowledge necessary for providing specific, actionable guidance on disclosure communication strategies~\cite{Paley2021,Olsen2022}. Expert-based assessment, though higher quality, cannot scale to meet the training needs of medical education programs seeking to provide frequent practice opportunities for all residents.

\subsection{Human-AI Collaboration for Healthcare Communication Training}
\label{subsec:related-ai-healthcare}

The integration of artificial intelligence into healthcare education has emerged as a promising approach for addressing scalability and personalization challenges in medical training. AI-assisted simulation systems have demonstrated effectiveness in clinical skill development, offering adaptive learning environments that respond to individual learner needs~\cite{zary2009web,cook2010robots}.

In the domain of medical communication training specifically, conversational AI systems have shown potential for supporting skill development in patient interactions. Chatbot-based training platforms enable students to practice clinical conversations with simulated patients, providing immediate feedback on communication choices~\cite{laranjo2018conversational,bickmore2018patient}. These systems offer advantages over traditional training methods including consistent availability, standardized scenarios, and objective assessment capabilities.

Recent advances in large language models have opened new possibilities for sophisticated medical communication training. Natural language processing technologies can now generate contextually appropriate patient responses, assess communication quality across multiple dimensions, and provide personalized feedback aligned with established communication frameworks~\cite{lee2022evaluating,bommasani2021opportunities}. Language models demonstrate particular promise for supporting complex communication scenarios where patient responses must adapt dynamically to learner communication choices.

Research in AI-supported medical education has identified key design principles for effective human-AI collaboration in training contexts. Successful systems balance AI automation with human oversight, provide transparent feedback mechanisms, and maintain learner agency in the learning process~\cite{roll2016evolution,luckin2016intelligence}. These principles are particularly important in healthcare training, where communication skills directly impact patient care outcomes.

However, existing AI applications in healthcare communication training have primarily focused on routine patient interactions rather than high-stakes scenarios like medical error disclosure. The emotional complexity, ethical dimensions, and institutional implications of disclosure conversations present unique challenges for AI-assisted training systems that have not been fully addressed in current literature.

\subsection{Design of Human-Language Model Interaction Systems}
\label{subsec:related-human-language-interaction}

The emergence of sophisticated language models has catalyzed development of interactive systems that enable humans to collaborate with AI for complex task completion. These human-language model interaction systems facilitate environments where users can leverage AI capabilities while maintaining control over task outcomes and decision-making processes~\cite{amershi2019guidelines,wang2021artificial}.

Interactive language model systems have demonstrated effectiveness across diverse application domains. In creative writing, systems like those developed by Clark et al.~\cite{clark2018creative} enable authors to collaborate with AI for story generation, providing suggestions and alternatives while preserving creative control. Programming applications such as GitHub Copilot exemplify how language models can assist in code generation while allowing developers to guide and refine AI outputs~\cite{chen2021evaluating}. Similarly, brainstorming and ideation tools leverage language models to stimulate creative thinking while maintaining human judgment in idea evaluation and development.

Key design principles for effective human-language model interaction have emerged from research across these domains. Successful systems provide users with meaningful control over AI behavior, offer transparent feedback about system capabilities and limitations, and enable iterative refinement of AI outputs~\cite{wilder2020learning,yang2020re}. These systems balance AI assistance with human expertise, ensuring that domain knowledge and contextual understanding guide interaction outcomes.

In educational contexts, human-language model interaction systems show particular promise for personalized learning applications. Language models can adapt to individual learning styles, provide customized explanations, and generate practice scenarios tailored to specific skill development needs~\cite{roll2021artificial}. The interactive nature of these systems enables real-time feedback and iterative skill refinement that traditional educational technologies cannot provide.

However, designing effective human-language model interaction systems for professional training presents unique challenges. Healthcare education contexts require systems that not only support skill development but also align with professional standards, ethical guidelines, and patient safety considerations~\cite{topol2019high}. The high-stakes nature of medical communication training demands particular attention to system reliability, feedback accuracy, and integration with existing educational frameworks.

Our work extends this foundation by investigating how human-language model interaction can specifically support medical error disclosure training, addressing the unique requirements of this critical healthcare communication scenario while leveraging established design principles for AI-assisted learning systems.
\section{Study Design}

Our motivation for building CandorMD came from three directions. First, prior work on communication training technologies (e.g., the VCA system and Grossnicklaus et al.~\cite{grossniklaus2025incorporating,white2022video}) had already shown what learners and educators want in these tools: real-time feedback, voice-based simulation, and designs that are both affordable and scalable. Second, our research team includes educators who have spent decades teaching error disclosure skills in clinical training programs, giving us practical insight into what works and where training often falls short. Finally, we built on broader line of work in improving communication, interpersonal, and therapeutic skills~\cite{lin2024imbue,sharma-etal-2023-cognitive}, which pointed to the importance of realistic practice and just-in-time feedback.  

We designed and iteratively developed a simulation \& feedback prototype, CandorMD. The system went through multiple informal testing rounds before we invited key informants to use it. We saw this not only as a chance to refine the system, but also as a way to bring diverse perspectives into shaping what effective training could look like.  

To further understand both current practices and opportunities for improvement, we conducted semi-structured interviews with 12 key informants across the disclosure ecosystem: practicing physicians, risk managers, communication researchers, and patient advocates. These informants shared their experiences with error disclosure and reflected on existing training approaches and their limitations. We also asked them to try the prototype and provide feedback on its design and potential pedagogical value.  

In the following sections, we describe the prototype design (Section~\ref{sec:demo_design}), interview procedure and analysis (Section~\ref{sec:interview}), and prototype evaluation (Section~\ref{sec:demo_interview}).

\subsection{Prototype Design}
\label{sec:demo_design}
Our prototype evolved through multiple iterative design cycles that integrated expertise in education, clinical communication, and human–computer interaction. \textbf{Importantly, this system was co-designed with domain experts who have decades of experience teaching medical error disclosure.} Across development, we incorporated successive rounds of feedback, including informal sessions with clinicians, to surface practical concerns and refine both content and interaction design. These iterations allowed us to embed design principles accumulated through prior research and professional practice, resulting in a prototype that was both grounded in theory and responsive to real-world training needs.
 
CandorMD is built around two core agents -- a \textbf{Patient Simulation Agent} and an \textbf{Evaluator Agent} -- that operate over a shared conversation but maintain distinct roles, information access, and internal reasoning. In what follows, we describe each agent's design, how they coordinate, and the principles that guide their interaction.


\begin{figure}[ht]
    \centering
    \includegraphics[width=\linewidth]{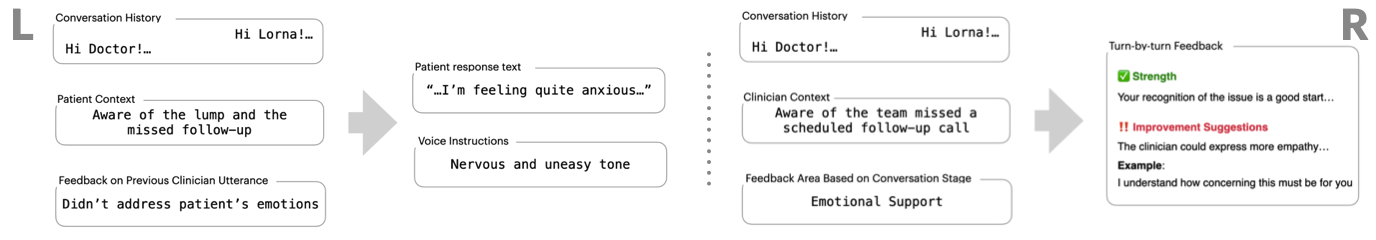}
    \caption{Simulation and feedback pipeline in CandorMD. In the simulation module (L), patient response text and voice instructions are simultaneously generated based on the conversation history, the patient’s knowledge of the error, and feedback on the clinician’s prior utterance. In the feedback module (R), turn-by-turn guidance is produced from the conversation history, clinician context, and the feedback area corresponding to the current conversation stage, highlighting both strengths and areas for improvement. See Section~\ref{sec:demo_design} for more information.}
    \label{fig:generation_combined}
\end{figure}

\subsubsection{Patient Simulation Agent}
 
Simply telling a language model to ``play the role of a patient'' often results in unrealistic behavior: the model tends to be overly agreeable, or audio models produce rigid emotional tones (e.g., always angry or neutral) without nuance. These limitations make practice feel less realistic and reduce pedagogical value. To address this, CandorMD's Patient Simulation Agent generates both the patient's textual response and detailed voice instructions (e.g., pacing, descriptive emotions) at the same time. This ensures the emotional delivery matches the patient's words. The generation of detailed voice instructions happens at every turn, allowing more dynamic, real-time, and fine-grained control of the patient's emotional changes throughout the conversation. We then use OpenAI's GPT-4o-mini-tts, a text-to-speech model that allows specification of text and emotional cues, to generate the simulated patient audio response.
 
The Patient Simulation Agent maintains its own internal state, consisting of three components:
 
\begin{itemize}
    \item \textbf{Information context:} The patient's evolving knowledge of the medical situation -- what they have been told, what they suspect, and what remains unknown to them. This is managed separately from the physician's knowledge to enforce a realistic information asymmetry.
    \item \textbf{Affective state:} A set of emotional dimensions (e.g., anxiety, anger, trust, confusion, grief) represented as continuous values that update each turn based on the agent's interpretation of the physician's utterance. These values drive the voice instruction generation, bridging internal state to the audio output.
    \item \textbf{Conversation memory:} A bounded window of recent conversational turns that provides the agent with dialogue context for generating coherent, contextually appropriate responses.
\end{itemize}

As shown in Figure~\ref{fig:generation_combined} (L), the agent takes three main inputs when generating a response: the conversation history, the patient's current information context, and the clinician's performance in the previous turn. The patient's knowledge shapes their mental state and influences what they might say next. We also incorporate clinician performance as an explicit input so that the agent's responses adapt more directly: when communication goes well, the patient becomes more receptive; when it does not, the patient may withdraw or push back. Taken together, these inputs drive both the patient's words and emotional tone, allowing behavior to shift with the flow of the conversation and making practice more realistic and useful. We use OpenAI's GPT-4o model to generate the responses; the prompts we used are in Appendix~\ref{app:prompts}.

\subsubsection{Clinician Evaluator Agent}
Training communication skills requires feedback that is both immediate and reflective. Turn-by-turn guidance can be valuable for shaping responses in the moment, but it also makes practice sessions longer and cannot capture broader patterns. For example, while an apology is often expected at some point in a disclosure, it is not appropriate at every turn—this type of feedback is only meaningful at the conversation level. To address both needs, CandorMD's Evaluator Agent provides two complementary forms of feedback: (1) utterance-level feedback during the simulated conversation, and (2) overall feedback at the end of the session. We included both in our prototype to examine their pedagogical value during interviews.
 
The Evaluator Agent operates with a distinct role and information access from the Patient Simulation Agent. As shown in Figure~\ref{fig:generation_combined} (R), to generate feedback, the agent analyzes the conversation history, the physician's knowledge of the situation, and the patient's most recent response. Because disclosure is inherently patient-centered, our evaluation framework emphasizes how well physicians respond to patient cues rather than simply delivering information. The agent operationalizes feedback by first detecting the conversational stage reflected in the patient's utterance and then mapping it to relevant evaluation goals. Conversations are rarely linear: multiple goals can arise simultaneously, and patients may shift between stages or return to earlier ones. The agent therefore adapts feedback dynamically, offering guidance that reflects the complexity and fluidity of real disclosure encounters. We use OpenAI's GPT-4o model to generate the feedback; the prompts we used are in Appendix~\ref{app:prompts}.
 
The Evaluator Agent maintains its own internal state, including running assessments across four feedback dimensions and a record of which conversational stages have been addressed. Building on prior work that collected expert annotations of VCA responses and ratings~\cite{white2022video}, we developed an evaluation framework to guide the agent's feedback. We refined and classified existing annotations into four categories, drawing on established frameworks for disclosure such as SPIKES~\cite{baile2000spikes}. These four metrics correspond to distinct conversational stages and feedback domains, as summarized in Table~\ref{tab:assessment_framework}. Some additional evaluation of the feedback generation is in Appendix~\ref{app:feedback_score}.

\begin{table}[h]
\centering
\resizebox{.45\textwidth}{!}{%
\begin{tabular}{@{}ll@{}}
\toprule
\textbf{Conversation Stage}   & \textbf{Feedback Area}        \\ \midrule
Information Seeking           & Acknowledgment \& Explanation \\
Emotional Expression          & Emotional Support             \\
Trust \& Accountability       & Trust \& Accountability       \\
Forward-looking Care Planning & Resolution                    \\ \bottomrule 
\\
\end{tabular}}
\caption{When generating feedback, the system first detects the conversational stage reflected in the patient’s response and then identifies the corresponding feedback area. This ensures a patient-centered communication flow by guiding feedback to address what is expressed in the patient’s response—rather than evaluating every possible dimension at once—thereby highlighting what is most relevant at that moment in the disclosure conversation.}
\label{tab:assessment_framework}
\end{table}



\subsubsection{Agent Coordination and Information Management}
 
A key design principle in CandorMD is how the two agents coordinate. Rather than operating as independent modules, the Patient Simulation Agent and Evaluator Agent are coupled through a shared conversation state but maintain separate information access and distinct reasoning processes. This coordination has two important properties.
 
\paragraph{Joint generation.} Simulated patient responses and feedback are generated together, as a multi-task problem. This keeps them aligned: the patient's reaction makes sense in relation to the feedback, and vice versa. In this way, the feedback is reinforced through the simulated interaction, giving clinicians a more lifelike training experience. Without this coupling, learners could face situations where the patient's response and the system's evaluation point in different directions, undermining the training.
 
\paragraph{Information asymmetry management.} The two agents have access to different information about the case. The Evaluator Agent has access to the full clinical context—including what the physician knows—in order to assess communication quality accurately. The Patient Simulation Agent, by contrast, has access only to what the patient would realistically know at each point in the conversation. This creates a natural information asymmetry that mirrors real disclosure encounters, where the physician knows more about the error than the patient.
 
Managing this asymmetry introduces a challenge: feedback is generated based on physician knowledge at a given turn, which can include information the patient has not yet been told. To avoid ``leaking'' physician knowledge prematurely into the patient's responses, we filter the Evaluator Agent's output into keywords (without sensitive details) before using it to guide the Patient Simulation Agent's next response. This ensures that the patient agent's behavior remains grounded in what the patient would plausibly know, while still allowing the evaluator's assessment to shape the overall trajectory of the conversation.

\subsection{Participant Recruitment}

Participants were recruited through professional networks and snowball sampling to ensure representation across diverse institutional contexts and roles. We purposively selected them to capture complementary perspectives on the system’s design and learning outcomes. All participants provided informed consent for participation, audio recording, and transcript extraction. Participants were compensated with \$75 gift cards of their choice. We recruited 12 participants from different institutions (Table \ref{tab:participants}).

We obtained approval from our institution's Institutional Review Board (IRB) for our study. Our institution requires all researchers who conduct human subjects research to complete human subjects protection training. The researchers who conducted this study were certified by the IRB.

\begin{table*}[t]
\centering
\small
\begin{tabular}{@{}p{0.23\textwidth}c p{0.68\textwidth}@{}}
\toprule
\textbf{Role of Participants} & \textbf{Number} & \textbf{Expertise Area} \\
\midrule

Practicing Clinicians (C) & 4 &
Clinicians experienced in conducting medical error disclosure conversations. Expertise includes clinical disclosure, patient communication, and specialty-specific challenges. \\

Risk Managers (RM) & 2 &
Hospital administrators and Communication \& Resolution Program (CRP) managers responsible for disclosure policies and physician support. Expertise includes institutional protocols, best practices, training coordination, and regulatory compliance. \\

Communication Researchers (CR) & 3 &
Specialists in healthcare communication and disclosure training. Expertise includes communication theory, training development, and physician education. \\

Patient Advocates (PA) & 3 &
Representatives working with patients and families affected by medical errors. Expertise includes patient perspectives, communication impact, and advocacy services. \\

\bottomrule
\end{tabular}

\caption{Roles, number of participants, and expertise areas of our interview participants. In Section~\ref{sec:findings}, we use the abbreviations to mark the participant roles in their quotes.}
\label{tab:participants}
\end{table*}

\subsection{Interview Process}
\label{sec:interview}
We conducted a one-hour interview with representatives from four distinct key informant groups, which were purposively selected to capture complementary perspectives on the system’s design and learning outcomes: clinicians (users), patients (beneficiaries), risk managers (organizational stakeholders), and patient–provider communication experts (evaluators). 
We began each interview by exploring the participant's background and relevant consent procedures.
The interviews consist of two parts: (a) semi-structured interviews that explore participants' insights on medical error disclosure, (b) prototype interaction and feedback.
Based on our designed interview protocol (Appendix~\ref{app:interview_protocol}), we aimed to explore four research questions, each adapted to the participant's role: 


\begin{itemize}
    \item \textbf{RQ1 -- Key Factors}: What interpersonal, organizational, and regulatory factors shape whether and how clinicians disclose medical errors?
    \item \textbf{RQ2 -- Current Practices}: How do clinicians currently prepare for disclosure conversations, and what are the current approaches that facilitate long-term learning?
    \item \textbf{RQ3 -- Challenges and Shortcomings}: What are the challenges associated with disclosing medical errors, and what are the shortcomings of current methods for learning and preparation?
    \item \textbf{RQ4 -- Technology-Supported Training}: How can technology support preparation and learning for disclosure, as explored through a prototype design probe?
\end{itemize}


\subsubsection{Prototype Interaction and Feedback}
To evaluate CandorMD's usability and gather expert feedback on system design, we collect in-situ and post-hoc feedback with key informants. The process is inspired by the think-aloud protocol~\cite{talkaloud-ericsson1998study, talkaloud-van1994think,talkaloud-medical-gegenfurtner2013transfer}.  We encouraged all participants to provide in-situ verbalization during their interaction with CandorMD, enabling real-time collection of user cognitive processes, identifying usability issues and design opportunities that may not emerge through post-hoc interviews alone~\cite{talkaloud-ericsson1998study}.  But since among the participants population, only clinicians are the target users, for patient advocates, risk managers, and communication experts, we provided more instructions during their interaction with our demo and focused on slightly different aspects when collecting feedback.


Each session lasted approximately 30 minutes and followed a standardized protocol. Sessions were conducted individually via video conference with screen sharing enabled to observe participant interactions with the system interface.

Prior to system interaction, participants received brief instructions on in-situ verbalizaiton, emphasizing the importance of verbalizing thoughts and reactions continuously throughout the session~\cite{nielsen1994usability}. Participants were assured that there were no correct or incorrect ways to use the system and that the goal was to understand their natural interaction patterns and reactions. 

\xhdr{System exploration and in-situ verbalization}: Participants were first shown the CandorMD landing page and asked to explore the system freely while verbalizing their initial impressions, expectations, and navigation decisions. Participants were then asked to complete one or more disclosure scenarios using the system, working through all steps from scenario selection, simulation, and feedback. They were encouraged to verbalize their thought processes, reactions to system responses, and assessment of system utility throughout the task.

\xhdr{Reflection and feedback}: Following task completion, participants engaged in a brief reflective discussion about their overall experience, potential improvements, and integration possibilities with current training approaches.

All sessions were video recorded with participant consent. The facilitator employed minimal intervention techniques, providing clarification only when participants experienced technical difficulties or extended periods of silence.

\subsection{Analysis Methodology: Thematic Analysis}

We employed systematic thematic analysis to extract insights from interview transcripts. Audio recordings were transcribed within Zoom and anonymized. We used Claude to clean the transcripts by removing disfluencies while preserving the original meaning of participants' responses, and spot-checked to double check. A comparison of before and after cleaning is in Appendix \ref{app:transcripts_clean}. Transcripts were manually segmented largely by interview questions to facilitate further thematic analysis. Transcripts were divided by conversational turns, with very long responses further broken down into multiple turns based on their meaning during coding.

Two researchers collaboratively developed an initial coding framework, analyzing 15 representative transcript segments for each research question, establishing a shared understanding of coding categories and application criteria. The codes for the material were developed inductively. Following framework development, both researchers independently coded all remaining transcript segments using the established framework while remaining open to emergent themes. 

Annotation disagreements were resolved through discussion and consensus, with a third researcher consulted when needed. Final coded segments were grouped thematically and analyzed to identify patterns across participant roles and convergent insights about current practice limitations.

\section{Study Findings}
\label{sec:findings}

\subsection{RQ1: Key factors in medical error disclosure conversations} 
We identified key themes that emerged when participants discussed the most important aspects and current gaps in medical error disclosure practice. These themes reveal both the foundational elements that facilitate effective disclosure and the complex systemic barriers that impede it.

\subsubsection{Trust as the Foundation of Disclosure}

Trust is the foundation of therapeutic relationships that either facilitates or hinders effective error disclosure conversations. It is often challenging to build and easily lost, especially in high-stakes situations where medical errors have occurred.  

Participants noted that trust may exist from prior relationships, but in acute care it often needs to be created under pressure and in emotionally charged contexts:  

\begin{displayquote}
\textit{Often when I'm talking to folks, they're people I know pretty well, so I think maybe that helps a little bit when I go in, because I think that you have a basic level of trust which is there already. If you see someone in the hospital, sometimes you have to develop that trust within a day in a really stressful situation when something has gone wrong. And that is a really challenging thing to do.} [C1]
\end{displayquote}

\begin{displayquote}
\textit{Sometimes it works, sometimes it doesn't. Families in the moment are often very emotional. They're processing grief, or they're processing fear. They're vocalizing their distrust in the system as a whole. You're not just having a factual, scientific discussion. You're really trying to help a human being feel like they're being heard, that they're important, and that you want to find the answer too, really badly.} [PA1]
\end{displayquote}

Clinicians also highlighted how well-intentioned but premature conclusions risk undermining trust:  

\begin{displayquote}
\textit{And then I'll go through the facts that we know. And what I'm really careful to do is talk about what is known and not to jump to conclusions, because often I find that as clinicians, we blame ourselves, we jump to conclusions, and never blame someone else. And I think that if you do that, and then it turns out what you say is actually not accurate later on, that can really result in a lack of trust.} [C1]
\end{displayquote}

At the same time, disclosure done well was seen as a way to restore relationships:  

\begin{displayquote}
\textit{From my point of view, ``disclosure'' is a weird term, because what it's really about is developing a relationship. When there's a medical error, there's a betrayal because something didn't go right. When there's a betrayal, the relationship is at risk. A disclosure done well rebuilds that trust because there's an honest response to what happened and a coming together to figure out what that means for all parties.} [PA3]
\end{displayquote}

\subsubsection{Patient-Centered and Empathic Disclosure}

Medical error disclosure conversations need to be patient-centered.  

From the very start, how a conversation unfolds depends strongly on what and how much the patient already knows. Clinicians prepare differently in these situations—sometimes it can feel like ``jumping into cold water.''  

\begin{displayquote}
\textit{It will depend on if the patient knows something has gone wrong or not, and that guides the language I use. [...] One of the challenges is it's almost like jumping into cold water--you just want to get it over with. But allowing space for the patient and family to talk is really important. They should probably be talking more than the physician.} [C1]
\end{displayquote}

Family members and caretakers are often part of the process, especially when patients are acutely ill or less verbal. Disclosure can range from small omissions to more serious errors, and often involves partners, guardians, or loved ones.  

\begin{displayquote}
\textit{Error disclosure can be as small as, ``We forgot to write the order,'' or as serious as missing an infection or adverse drug reaction. In those cases, it often needs to be discussed with family, partners, or guardians.} [C4]
\end{displayquote}

These conversations also vary depending on patient response. For some, acknowledgment is enough; for others, disclosure can trigger distress or distrust, requiring multiple follow-ups.  

\begin{displayquote}
\textit{Sometimes patients say, ``Fine, thanks for letting me know,'' and it's done. Other patients are really distressed and feel distrustful of the team or of medicine as a whole. [...] Sometimes it gets revisited several times to help the patient process.} [C4]
\end{displayquote}

Finally, participants emphasized that disclosure is not a one-sided act. Historically too physician-centric, good communication requires patients' voices, preferences, and concerns to shape the interaction.  

\begin{displayquote}
\textit{When I started in this field, the research was so physician-centric. [...] In reality, good communication depends on both parties, because it takes both to create an interaction.} [CR1]
\end{displayquote}

\begin{displayquote}
\textit{My focus was [...] on ensuring that patients express their preferences, interests, and concerns so they and doctors can reach shared understanding and decisions.} [CR1]
\end{displayquote}

\subsubsection{Balancing Timeliness and Ongoing Dialogue}
Participants described timeliness as a central principle of effective error disclosure. Conversations should begin promptly after an error is identified, signaling accountability and responsiveness. At the same time, disclosure was framed as an ongoing process rather than a single exchange. Especially in high-severity cases, patients and families revisit the conversation with new questions, emotions, and needs. Effective disclosure thus requires balancing immediacy with continuity, sustaining dialogue over time to support trust and understanding.

Error disclosure should occur promptly after discovery and continue as an ongoing process rather than a single event. Participants emphasized the importance of immediate action, highlighting that conversations often happen very quickly and cannot wait for routine visits:  

\begin{displayquote}
\textit{Because typically, if I know something that's gone wrong, I won't wait for a scheduled visit, I'll call them and talk about them. I don't want to let time pass.} [C1]
\end{displayquote}

\begin{displayquote}
\textit{There isn't a lot of time if you have to go in right away to talk to somebody. You need somebody on the ground immediately to help process a little bit.} [PA3]
\end{displayquote}

At the same time, participants stressed that disclosure is not resolved in a single exchange. What begins as an urgent conversation must continue over time. Disclosure is rarely a ``one-and-done'' event especially for events with high severity:  

\begin{displayquote}
\textit{The first conversation should happen quickly after a medical error, but there will be subsequent conversations. [...] These conversations can transpire across days, weeks, months, or even years, depending on how egregious the harm is.} [PA1]
\end{displayquote}

\begin{displayquote}
\textit{One potential problem with thinking about disclosure is that it might not be one conversation. Somebody comes in to tell me about what happened with my daughter, we talk, but I go away and have all kinds of questions in the aftermath.} [PA3]
\end{displayquote}

\subsubsection{Ensuring transparency while addressing uncertainty}  
Transparency in disclosure is both an ethical obligation and a relational practice. Participants described four key dimensions. First, transparency is a matter of respect: patients deserve to know what happened so they can make informed decisions. Second, transparency requires balance—sharing what is known, acknowledging what is not, and avoiding speculation. Third, it involves admitting uncertainty rather than offering premature or inaccurate answers. Finally, systemic failures in transparency can create compounding harm, where the absence of honesty after an error adds new injuries to the original event.  

Participants emphasized that openness is not only an ethical obligation but also a matter of respect for patients:  

\begin{displayquote}
\textit{When you're talking to someone about something that's gone wrong in their healthcare, full transparency of the facts you know at the time is essential. For patients to make informed decisions going forward, they have to know what has happened in the past.} [C1]
\end{displayquote}

Maintaining transparency requires careful balance--distinguishing between what is known and what remains uncertain, while resisting the temptation to speculate:  

\begin{displayquote}
\textit{I'm really careful to talk about what is known and not to jump to conclusions. [...] Part of that is being upfront that there are things we don't know right now, but we are doing our best to find out and will update as it goes along.} [C1]
\end{displayquote}

Several participants noted that honesty also means explicitly acknowledging uncertainty rather than offering incomplete or inaccurate answers:  

\begin{displayquote}
\textit{I don't want to tell you something that's not accurate. If I find out later it isn't the case, that's harder to work through than simply saying, ``We're working as hard as we can, and we'll keep you updated.''} [RM1]
\end{displayquote}

Finally, participants highlighted how systemic failures in transparency can compound harm. Lack of honesty after an error creates compounding harm that is sometimes worse than the first:  

\begin{displayquote}
\textit{There's a call for honesty, transparency, and accountability in the aftermath of medical harm.} [PA3]
\end{displayquote}

\begin{displayquote}
\textit{While the first harm was unintentional, all the subsequent harms were intentional when you're not honest, transparent, and having those conversations.} [PA1]
\end{displayquote}

\subsubsection{Apology}  
Participants emphasized the importance of apology in error disclosure. While there is a common perception that clinicians are advised to avoid apologizing to reduce liability risk, participants described apology instead as an essential expression of empathy, respect, and accountability.

Apologies were framed first and foremost as empathic acts that acknowledge the patient's experience:  

\begin{displayquote}
\textit{I am so sorry that this happened to you.} [C1]
\end{displayquote}

\begin{displayquote}
\textit{This is not at all what we intended to happen to your medical care. I am so sorry that this happened.} [C1]
\end{displayquote}

At the same time, participants cautioned against a premature ``fault apology'' before the full circumstances are clear:  

\begin{displayquote}
\textit{I don't do that initial conversation as a fault apology—saying, ``you were harmed because I did this''—because often we don't know everything that has happened at that point. That kind of apology is more appropriate after the facts are established.} [C1]
\end{displayquote}

Finally, participants reflected on the evolution of their practices, moving away from avoidance toward transparency and accountability:  

\begin{displayquote}
\textit{Before I learned best practices, I sometimes glossed over errors with patients. As I learned better ways, I now champion what we should be doing: being transparent about what happened, apologizing, and submitting the event to the reporting system.} [C3]
\end{displayquote}

\subsection{RQ2: Current approaches for disclosure preparation and long term learning} 

Participants described preparation ranging from brief, ad hoc efforts before conversations to longer--term learning through training and professional experience. Across accounts, current methods were viewed as fragmented, inconsistent, and inadequate for the complexity of medical error disclosure. Preparation and learning opportunities often depend more on individual initiative or workplace culture than systematic training.

\subsubsection{Preparation and Long-Term Learning}  
Participants described a wide range of current approaches to preparing for disclosure conversations, noting that most remain informal, inconsistent, and highly dependent on individual experiences. Traditional medical education often relied on observation and imitation, encapsulated by the adage ``see one, do one, teach one.'' While valuable, participants emphasized that passive observation alone is inadequate for such complex and emotionally charged interactions:  

\xhdr{Preparation Before Conversations}  
Preparation for disclosure often happens ad hoc, in the moments leading up to the conversation. Participants described quick pre-meetings or one-on-one coaching to align messaging and anticipate likely questions. While helpful, these interactions are typically brief and lack the depth of structured rehearsal:  

\begin{displayquote}
\textit{For the initial disclosure, one of us calls the person who will lead and goes through it with them. The final disclosure is more formal, with the whole team present to share what we learned from the review and what we will do to prevent recurrence.} [PA3]
\end{displayquote}

\begin{displayquote}
\textit{Before difficult conversations like delivering a bad diagnosis, we’ll often sit down with the trainee and ask, ``What are you planning to say?’’ Then we reframe it together and go in as a team. But it’s usually three minutes of coaching, not twenty minutes of practice.} [C4]
\end{displayquote}

\begin{displayquote}
\textit{For goals-of-care conversations, we always have a pre-planning meeting and a post-meeting reflection. That should be happening for disclosure too, but I don’t know that it is.} [C1]
\end{displayquote}

Despite the current situation, participants emphasized that desired preparation should go beyond observation and ad hoc learning. They pointed to the need for structured models, repeated practice, and actionable feedback as essential for building competence in disclosure conversations.

\begin{displayquote}
\textit{As opposed to just winging it. You need to be clear, have a model to follow, and then get an opportunity to practice and receive feedback. That’s how people get better.} [CR2]
\end{displayquote}

\xhdr{Long-Term Learning and Skill Development}  
Beyond immediate preparation, participants reflected on how disclosure skills are (or are not) developed over the course of medical training. Historically, communication learning relied heavily on observation and imitation, which some participants felt was inadequate for such sensitive encounters:  

\begin{displayquote}
\textit{You would see how an attending communicated with a patient. The old idea was ``see one, do one, teach one.'' But for skills that involve another person’s emotions, passive learning isn't enough.} [RM1]
\end{displayquote}

\begin{displayquote}
\textit{During residency, I learned tools like the SPIKES acronym—setting, perception, information, knowledge, emotions, and strategy. [...]We practiced with role-play, [...]but most training was through real-life observation. [...]The practice was a small percentage, but very helpful.} [C2]
\end{displayquote}

Simulation frameworks, role-playing, and structured tools have been introduced, but remain sporadic and unevenly applied. Importantly, participants stressed that without consistent assessment or reinforcement, communication training is often deprioritized:  

\begin{displayquote}
\textit{Every medical school and residency program has their own program. But if we don't assess it, they don’t think it's important. When exams like Step 2CS were dropped, schools started pulling back on teaching this, because they respect what you inspect.} [PA3]
\end{displayquote}

\begin{displayquote}
\textit{If you think a vignette is too hard in a simulation where it's not you, how do you think it's going to be in real life? That gap shows how unprepared people are when training doesn't reflect reality.} [CR2]
\end{displayquote}

Participants also emphasized that long-term learning often comes from lived experience. The way a clinician’s first errors are handled—whether with respect and support, or with criticism—can shape their future willingness to engage openly in disclosure:  

\begin{displayquote}
\textit{How an error I was involved in was handled shaped how I approach disclosure. Because it was done respectfully and professionally, I became someone who now champions best practices. In another setting, being disrespected could easily have pushed me away from disclosure altogether.} [C3]
\end{displayquote}

\begin{displayquote}
\textit{By doing it repeatedly, because we make errors--we're human and we make mistakes. Early in training there's a sense you're not allowed to make mistakes, but that's not true. What matters is what you do with them.} [C4]
\end{displayquote}

\subsection{RQ3: Challenges, Gaps, and Identified Opportunities}

Participants described medical error disclosure as uniquely challenging, requiring clinicians to manage their own emotions while responding to the complex reactions of patients and families. These conversations demand skills beyond routine practice, yet training and feedback are often inconsistent or absent. Improving disclosure thus requires both cultural change and opportunities for specific, timely, and emotionally attuned feedback, potentially supported by simulation and AI tools.

\subsubsection{Clinicians' Own Emotional Regulation}  
Participants described clinicians own emotional regulation as one of the most difficult aspects of medical error disclosure. Even before responding to patients' or families' reactions, clinicians must contend with the internal weight of the error itself. Feelings of shame, guilt, anxiety, and fear of reputational or legal consequences were described as profound barriers to communication.  

Clinicians emphasized how personally devastating errors can feel, even when their role was indirect:  

\begin{displayquote}
\textit{Everyone makes mistakes, and nobody likes to look incompetent. For clinicians, a serious mistake is a horrific experience. Even if they had a secondary role, they still experience emotions alongside the patient's.} [CR3]
\end{displayquote}

Participants also noted that fear of lawsuits and reputational harm further compounds this challenge:  

\begin{displayquote}
\textit{Historically, there has been fear of lawsuits and malpractice. There is also fear of reputational damage—not just public exposure, but colleagues thinking you are not competent. Nobody wants that.} [CR3]
\end{displayquote}

These emotions were seen as directly interfering with the ability to carry out communication skills effectively. At the same time, emotional regulation itself was described as a skill that can be strengthened through deliberate practice and feedback:  

\begin{displayquote}
\textit{Clinicians feel embarrassment, shame, and guilt. These emotions can get in the way of communicating well.}  [RM2]
\end{displayquote}

Disclosure was contrasted with other high-stakes conversations, such as goals-of-care discussions to describe the unique challenges that comes with error disclosures:  

\begin{displayquote}
\textit{In a goals-of-care conversation, you're not worried you'll get sued, and it's not you saying, ``I made a mistake.'' Disclosure is different—it feels like saying, ``I failed you.'' That is a hugely different thing.} [C1]
\end{displayquote}

\subsubsection{Responding to Patients' or Families' Emotions}  
Participants identified responding to the strong emotions of patients and families as one of the most challenging aspects of disclosure that rarely get practiced. Anger, grief, shock, or silence were described as difficult to respond to in ways that feel authentic and supportive. Clinicians must acknowledge these emotions while maintaining composure and sustaining the relationship.  

Several participants emphasized that patients' or families' anger is often the most difficult response to navigate:  

\begin{displayquote}
\textit{The emotion is often the hardest part. Patients may lash out: ``How could you do this to me? I'm going to sue you. I'm going to bring this building to its knees.'' Those are difficult things to respond to in a productive manner. Doctors are humans too, and it wouldn't be hard for a response to come across poorly.} [RM1]
\end{displayquote}

Participants also stressed that emotions are rarely singular---patients and families often experience a complex mix that can include silence, guilt, or disbelief as much as anger:  

\begin{displayquote}
\textit{The number one emotion patients and families usually feel is guilt—guilt that they didn't do more, or that they let this happen. Some are stunned, others break down and cry, some get angry. Providers need to recognize the full range rather than assuming anger alone.} [PA1]
\end{displayquote}

\begin{displayquote}
\textit{Clinicians often go in assuming patients will be angry, putting on a suit of armor before saying a word. But many patients actually feel guilt, sadness, or shock. If you expect anger, you can unintentionally shape the entire tone of the conversation.} [PA1]
\end{displayquote}

\begin{displayquote}
\textit{What if the patient shuts down and says nothing? They may just be trying to digest, or breaking down in silence. Providers need to learn to sit with that, rather than rushing to fill the space.} [PA1]
\end{displayquote}

Authentic acknowledgment was seen as central. Clinicians must recognize emotions directly and respond in ways that show listening and care:  

\begin{displayquote}
\textit{If somebody is really upset, you have to respond authentically and let them know you hear them. You can't fake that.} [CR2]
\end{displayquote}

\begin{displayquote}
\textit{Giving space for the patient to express emotion—calling it out by saying, ``It sounds like this is really stressful or anxiety-provoking. Am I right?''—helps patients feel understood.} [C1]
\end{displayquote}

\subsubsection{Communication Skills}  
Participants emphasized that error disclosure requires a distinct set of communication skills that go beyond routine clinical interactions or other difficult communication situations. These include and are not limited to, apologizing in a way that conveys sincerity, translating complex medical information into understandable language, adapting to patient responses, and communicating with cultural sensitivity.  

One participant noted that even in routine care, many clinicians struggle to communicate clearly, making disclosure conversations especially fraught:  

\begin{displayquote}
\textit{Clinicians have a very poor track record on communicating with patients. As a family physician, it was common for patients to come to me just to find out what another doctor had told them because communication was so poor. Some clinicians are great at making patients feel heard, but the majority aren't—and it's only harder when the topic is medical error.} [PA2]
\end{displayquote}

Several participants stressed the importance of beginning with an appropriate apology and a basic overview, while committing to follow-up rather than overwhelming patients with premature detail:  

\begin{displayquote}
\textit{Up front you have to take some responsibility—``I'm sorry; this was not what we intended.'' Start by setting the right tone, give a basic overview, pledge to look into it, and come back once you've investigated.} [C3]
\end{displayquote}

\begin{displayquote}
\textit{The first phase is apologizing and accepting that something wasn't done correctly, then pledging to find out more and report it. The second phase is broader, often involving other team members to avoid misplaced blame and to show alignment across providers.} [C3]
\end{displayquote}

\begin{displayquote}
\textit{Transparency and honesty matter most. Don't beat around the bush or shift blame. A genuine, responsible apology is far more effective than one that feels required.} [C4]
\end{displayquote}

Participants also described the challenge of translating complex system dynamics and medical jargon into patient-understandable terms while avoiding both oversimplification and unnecessary alarm:  

\begin{displayquote}
\textit{You may understand the error from a system perspective, but patients don't. You have to distill what happened in a way that is fair and true, without overwhelming them or making them conclude the whole system is a disaster.} [C3]
\end{displayquote}

\begin{displayquote}
\textit{You want to be transparent, and you also don't want to overwhelm the patient with information far outside their knowledge. There's a need to translate complex facts into something easier to understand, while maintaining trust in the relationship.} [C3]
\end{displayquote}

Beyond factual clarity, communication skills include attending to emotions through silence, acknowledgment, and flexibility:  

\begin{displayquote}
\textit{Some strategies involve silence, which can be powerful. One of the most succinct findings in communication research is that patients in distress want to feel heard and understood. Listening is as important as speaking.} [CR1]
\end{displayquote}

Finally, participants underscored that communication must be adapted to different cultural and interpersonal contexts:  

\begin{displayquote}
\textit{What might be fine to say to one patient could offend another. In one disclosure conversation it was important, culturally, for the entire family to be present because decisions were made by consensus. Having a skilled communicator with cultural awareness was essential.} [RM1]
\end{displayquote}

\subsubsection{Current Feedback Mechanisms Are Insufficient}  

Participants highlighted the inadequacies of existing training and feedback systems for medical error disclosure. Structured reflection and coaching are rare in this context, leaving most clinicians with little systematic guidance and forcing them to rely on personal style or patient reactions.  

Several noted that traditional methods of assessment and training do not translate well to the interpersonal demands of disclosure, and that medical education has historically neglected communication altogether:  

\begin{displayquote}
\textit{When I started, about 30 years ago, we used multiple-choice tests. We gave scenarios and asked people to pick the best option. That didn't work well for communication. Later we experimented with standardized patients and even crowdsourced ratings, but feedback was inconsistent and not individualized.} [CR3]
\end{displayquote}

\begin{displayquote}
\textit{Nothing in medical school, and even through years in practice, this was not something anybody was showing me or teaching me. It came down to what kind of person you are. Most clinicians didn't get into medical school because of their interpersonal skills, and they're just on their own in this.} [RM1]
\end{displayquote}

Participants emphasized that even when feedback is present in other kinds of difficult conversations, such as goals-of-care discussions, these mechanisms rarely extend to disclosure:  

\begin{displayquote}
\textit{For goals-of-care conversations, we always have a pre-planning meeting and a post-meeting reflection. That's how we teach: talk about what went well, what was challenging, and what could change. That should be happening for disclosure too, but I don't know that it is.} [C4]
\end{displayquote}

In practice, many trainees receive no direct feedback at all. When attending clinican are absent, the only feedback comes from patient reactions—an approach participants viewed as inadequate and costly:  

\begin{displayquote}
\textit{In real life, the attending has to be there to hear the conversation, which isn't always the case. Sometimes the trainee gets zero feedback other than what they glean from the patient's reactions. Even when attendings are present, it's usually quick coaching--three minutes before going in-rather than sustained guidance.} [C3]
\end{displayquote}

Overall, participants described feedback and training in this domain as inconsistent, superficial, and insufficient for the stakes of disclosure conversations.

\subsubsection{Creating a Culture for Feedback}  
Participants emphasized that improving feedback on disclosure conversations requires more than technical fixes--it demands cultural transformation. Because communication is deeply personal, clinicians may resist feedback or fail to recognize its importance. A culture that normalizes feedback as supportive rather than judgmental is essential for growth.  

\begin{displayquote}
\textit{When you start talking about communication skills, it's getting personal. I have yet to meet a person in healthcare who admits they have poor communication skills. Some think they could improve, but nobody thinks they are horrible at it. So feedback needs to be actionable and sensitive—more positive than judgmental—because you are giving feedback on something that feels like a personal trait.} [CR2]
\end{displayquote}

Participants also highlighted that openness to feedback is shaped by mindset. Those who view disclosure as inherently difficult are more likely to engage and learn, while those who assume they already ``have it handled'' often fail to grow:  

\begin{displayquote}
\textit{People who think a situation is easy are actually less likely to learn from feedback, because they assume they already know how to handle it. By contrast, those who recognize the difficulty of these conversations are more receptive and improve more over time.} [CR2]
\end{displayquote}

\subsubsection{Desired Feedback Opportunities}  
Participants outlined specific ways feedback could be improved to better prepare clinicians for error disclosure. They emphasized that effective feedback must go beyond generic advice to be personalized, actionable, and attentive to the full complexity of these conversations.  

First, participants called for feedback that is concrete and immediately applicable:  

\begin{displayquote}
\textit{The feedback needs to be actionable. If you just say something like ``be more compassionate,'' what does that mean? How can I do that? Personalized, specific guidance is far more useful than generic comments.} [CR2]
\end{displayquote}

From a learning theory perspective, humans acquire behaviors most effectively when actions are paired with immediate outcomes. Delayed feedback not only diminishes its educational value but also increases cognitive load, as learners must retrieve the original context or risk ignoring feedback altogether~\cite{hattie2007power}. Participants echoed this concern:

\begin{displayquote}
\textit{Feedback can be personalized and almost immediate, whereas before there was a two-week lag. That makes a big difference.} [CR2]
\end{displayquote}

Participants also noted that effective feedback must help clinicians navigate the delicate balance between centering the patient and authentically showing up as human beings themselves:  

\begin{displayquote}
\textit{Sometimes we teach clinicians to just focus on the patient and they'll be okay. But a real conversation means they have to be present too. If I know the physician is devastated, that matters to me. It shows they care. Erasing the physician from the conversation is not honest.} [PA3]
\end{displayquote}

Finally, several participants pointed to the potential of simulation and AI tools for practicing emotionally charged scenarios. These tools could provide realistic opportunities to respond to anger, grief, or silence in ways that sustain presence and de-escalate tension:  

\begin{displayquote}
\textit{Our culture doesn't deal well with anger. Patients are volatile, and raw emotion can set you off. An AI tool could simulate those moments, helping clinicians practice staying present so that tension dissipates and the real conversation can emerge.} [PA3]
\end{displayquote}

Taken together, participants envisioned feedback that is specific, timely, emotionally attuned, and realistic, preparing clinicians to engage authentically in one of the most challenging conversations in healthcare.


\subsection{RQ4: Prototype Interaction and Evaluation of CandorMD}
\label{sec:demo_interview}

We analyzed the in-situ verbalizations and post-hoc reflections using inductive coding and organized findings into four areas: simulation fidelity (\S\ref{results:simulation}), feedback quality (\S\ref{results:feedback}), usability insights (\S\ref{results:usability}), and integration \& use cases (\S\ref{results:integration}).

\subsubsection{Simulation Fidelity}
We summarize participants' comments into three dimensions—emotions, factual grounding, and conversation flow--to reflect their evaluation of the level of realism that CandorMD conveys.  
\label{results:simulation}

\xhdr{Emotions} Participants generally found affect plausible, especially anger escalation when patient's needs were not met. One clinician noted the patient's ``anger is very good,'' yet some segments felt ``a little artificial'' when prosody did not match content. Several emphasized vocal nuance (pace, pitch, ``catch'' in the voice) and dynamic change across turns for realism.  
\begin{displayquote}
\textit{The program ratchets emotion up if needs aren't met[...] It doesn't let you off the hook.} [CR2]

\textit{The voice sounded realistic, though a little formal[...] speaking too fast.} [C2]
\end{displayquote}
\xhdr{Factual grounding} Participants valued transparency without premature specifics. They cautioned that the system sometimes nudged users to ``do it all in one go'' (full explanation, apology, systems fixes), which risks over-promising and inventing details not yet known. 
\begin{displayquote}
\textit{Say `I don't know yet, but I'm looking into it' rather than assume facts.}  [CR3]
\end{displayquote}
\xhdr{Conversation flow} Clinicians contrasted sentence-like turns with natural speech and asked for pacing to establish rapport before disclosing details, plus options to restart or course-correct mid-case. Views on system latency varied: a ``thinking'' indicator helped some, while others wanted faster turn-taking.  
\begin{displayquote}
\textit{People don't talk in perfect sentences[...] that's how people talk.}  [CR2]
\end{displayquote}

\subsubsection{Feedback Quality}
\label{results:feedback}
Participants shared their thoughts on immediate, turn-by-turn feedback and end-of-session summaries; many preferred having both. They favored concrete exemplars over generic advice, avoiding medical jargon, and role-aware accountability.  

\xhdr{Actionable exemplars} Examples (``You could say...'') were preferred to rubric-only comments; turn-by-turn guidance helped mid-conversation adjustment, while summaries supported reflection.

\xhdr{Balance \& tone} Participants recommended separating strengths and opportunities for improvements visually to avoid discouragement; some found per-turn ``nitpicks'' overwhelming.

\xhdr{Accountability nuance} Encouraging apology and ownership was valued, but participants warned against assigning individual blame or presuming certainty on accountability before reviews.

\begin{displayquote}
\textit{Real-time feedback is impressive[...] the summary helps reflect.} [CR3]

\textit{Small errors, like calling a patient by the wrong name, can really damage trust. The tool captured that well.} [CR1]

\textit{Numbers are fine, but concrete strengths and 'things to try' are most useful.} [C3]

\textit{I also think the summary feedback, probably at the very end, would allow more back and forth to allow you, as you would in real life. To hit all these points, but not just to dump it all at once.} [C1]

\textit{I agree with the feedback. I didn't expressly acknowledge the patient's anxiety, and I should have. I should have used that word, said ``I understand that's what you're experiencing,” because that helps the patient feel heard.} [RM1]
\end{displayquote}

\subsubsection{Usability Insights}
\label{results:usability}
Participants found the interface straightforward once started (e.g., ``It's very easy to navigate''), but flagged wording and control affordances (e.g., ``End conversation'' vs. ``Click to stop'') and requested clearer microphone permissions and restart/course-correction controls. Minor content/UI details were noted.  

\xhdr{Controls \& affordances} Participants suggested clearer differentiation between stopping and ending a case, as well as the addition of a restart-from-here option to support mid-conversation course correction. 

\xhdr{Audio realism} Several participants emphasized the importance of calibrating voice speed and intonation to better match conversational norms. They also noted that allowing natural pauses would increase the sense of realism.  

\xhdr{Health literacy} Participants preferred the use of plain language over technical jargon, highlighting the need for accessible communication. They further recommended tailoring exemplar responses to align with patients’ levels of understanding.  

\begin{displayquote}
\textit{Differentiate 'in conversation' from 'click to stop'--a forcing function or stronger visual contrast would help.} [PA1]

\textit{It's better with a face, but photos can prime; consider neutrality.} [CR3]
\end{displayquote}

\subsubsection{Integration and Use Cases}
\label{results:integration}
Participants envisioned two primary adoption modes: (1) \textit{just-in-time rehearsal} before difficult disclosures, and (2) \textit{curricular use} in medical school, residency, or on the job training. Flexibility to load user-defined scenarios and vary patient personas (angry, tearful, withdrawn, accepting) was suggested by two participants and viewed as essential.  
\begin{displayquote}
\textit{This is an order of magnitude better than vignette tools[...] excellent for helping a physician prepare.} [C3]

\textit{I could see this being helpful for someone who wants to actually input their scenario, for example, so they can kind of practice in that particular situation.} [C1]

\textit{Overall, this seems like a powerful tool. A library of simulations would be useful. If I had to disclose an error or give a bad diagnosis, I could practice a matching case beforehand. That would be very helpful.} [C2]
\end{displayquote}

Overall, clinicians described the patient simulation as both realistic and valuable, noting that its emotional trajectories, handling of factual uncertainty, and conversational pacing aligned closely with real practice. At the same time, they emphasized opportunities for improvement, including reducing feedback latency, incorporating more natural spoken language, and refining speech dynamics to further enhance authenticity and learning value.

\section{Discussion}
\label{sec:discussion}


Our interviews with physicians, risk managers, communication experts, and patient advocates reinforced that effective error disclosure are challenging. Existing training modalities address parts of the problem but are fragmented, inconsistently applied, and rarely provide longitudinal practice or individualized coaching. As a result, clinicians are often under-prepared for the emotional volatility, uncertainty, and evolving relational dynamics of real-life disclosures.

Our CandorMD prototype aims to address several of these barriers. First, the simulation enables repeated practice with emotionally charged responses (e.g., anger, silence, grief), allowing clinicians to rehearse before high-stakes encounters. Second, just-in-time feedback tackles a widely reported deficit by offering immediate, actionable, and patient-appropriate guidance. Unlike generic critiques used in prior work~\cite{white2022video}, CandorMD emphasizes concrete exemplars and distinguishes between strengths and areas for improvement, supporting reflection without discouragement. 

Together with the interview findings, we organize the future design discussion around two broad themes: technical design implications and challenges that extend beyond technical solutions. This division is not absolute, as advances in one domain can inform and support the other.







\subsection{Technical Design Implications}

\xhdr{Broader patient personas} CandorMD allows participants to describe their own cases beyond those provided, enabling diverse patient and scenario adaptation for practice and feedback. For pedagogical purposes, educators can input cases based on specialty, specific situations, and patient profiles. CandorMD also supports conversations directly with patients as well as with their family members or caregivers. These features were well received by participants. Future improvements could include adding more diverse patient personas—such as patients with limited English proficiency, families making decisions collectively, or individuals with varying levels of medical knowledge--to better mirror the challenges of real-world encounters. Although the current system offers such flexibility, these features require more careful testing before integration into educational use. Patient proficiency may pose a unique challenge for AI systems, which tend to generate overly fluent content. Expanding this breadth would help clinicians practice adapting communication to varied backgrounds and expectations.

\xhdr{Expanded emotional range}  
Participants consistently emphasized that responding to patient and family emotions is one of the most challenging aspects of disclosure. While anger and grief are frequently anticipated, participants described a broader and more nuanced spectrum that includes silence, guilt, disbelief, and resignation. CandorMD supports flexible emotional responses that adapt dynamically to the unfolding conversation, enabling clinicians to practice navigating both overt and subtle expressions. Future systems should expand this capacity by incorporating a wider range of emotional categories and intensities, while also giving participants agency to select the types of responses they wish to practice. Silence, in particular, emerged as a critical but often overlooked reaction; unlike anger or sadness—where clinicians are often mentally prepared—quiet patients demand different communicative strategies. As AI models advance in representing fine-grained emotional states, training systems can leverage this granularity to help clinicians rehearse responses to diverse, less predictable cues. Through repeated practice across varied scenarios, clinicians can strengthen their ability to acknowledge emotions authentically and respond with greater flexibility.

\xhdr{Agency on Feedback Format} 
CandorMD supports two modes of feedback: turn-by-turn and overall summaries delivered at the end of the conversation. Participants expressed mixed preferences. Some favored turn-by-turn feedback, highlighting its immediacy and ability to pinpoint specific conversational moments for improvement. Others preferred overall feedback, noting that real-time interruptions could disrupt the natural flow of conversation and that reflective synthesis afterward felt more authentic. Several participants saw value in combining both approaches--using turn-by-turn feedback to sharpen specific skills and overall summaries to consolidate lessons. These divergent perspectives underscore that no single format is universally optimal. Training systems should therefore give participants agency to decide when and how feedback is delivered, accommodating different learning styles.

\xhdr{Longitudinal support} Disclosure is rarely a ``one-and-done'' event. Participants emphasized that it unfolds across days, weeks, or even months, often requiring multiple follow-ups as patients and families process new information and emotions. CandorMD takes a first step by allowing participants to indicate whether the reason for the error is clear or still under investigation, and to edit the patient’s current understanding of the situation. Future systems could extend this by supporting evolving case knowledge and tracking how disclosure stages progress over time. Such tools might help clinicians plan follow-up conversations, monitor commitments made during earlier discussions, and reflect on how dialogue develops across encounters. By shifting from isolated conversations to longitudinal preparation and support, training systems could better align with realistic disclosure practice.

\xhdr{Just-in-time preparation vs long-term learning} Participants envisioned multiple use cases for CandorMD. On one hand, they saw value in just-in-time preparation, where clinicians could rehearse a disclosure conversation immediately before it occurs--particularly important given the shortage of experienced communicators or senior colleagues available for real-time coaching. At the same time, some cautioned that using clinicians’ own cases for practice raises legal and privacy concerns that must be carefully considered before deploying such tools in practice. On the other hand, participants also emphasized the value of CandorMD in educational contexts, such as residency training or workplace learning. In these settings, instructors could adapt the system by inputting specialty-specific cases, tailoring scenarios to reflect diverse clinical domains and training needs.

\subsection{Remaining Challenges Beyond Technical Solutions} 
Important needs remain outside the scope of technical interventions. While participants across roles valued CandorMD, they consistently underscored structural, cultural, and emotional barriers that technology alone cannot resolve. 

\xhdr{Systemic trust repair} After medical harm, patients evaluate trust not only through the disclosure itself but through system-level actions such as transparent investigation, consistent follow-up, and continuity of care. Patient advocates warned that disclosure without organizational accountability may compound harm. Several participants highlighted the value of a dedicated staff contact--beyond the attending physician--to help families navigate these processes, showing that effective support often extends beyond communication alone. 

\xhdr{A culture for learning and feedback} Even the best AI guidance will falter in environments that stigmatize error or discourage vulnerability. Communication experts observed that clinicians often resist feedback because it feels like a judgment of personal character: ``When you start talking about communication skills, it's getting personal. Nobody thinks they’re horrible at it.'' Participants noted that medical culture frequently treats mistakes as unacceptable, creating barriers to open discussion. Constructively, feedback framed as growth-oriented and supported by institutional leaders was seen as essential for making communication training safe and effective. 

\xhdr{Cultural and institutional constraints} Liability concerns, apology policies, and CRP protocols shape what clinicians are permitted to disclose, and when. Risk managers explained that even skilled communicators are often constrained by institutional rules: disclosure ``isn't just about what the clinician wants to say -- it's what the institution allows them to say, when, and how.'' Training tools can help clinicians with phrasing, but policy and culture strongly influence how disclosure is carried out in practice.  

Taken together, these findings suggest that AI can be a valuable component of disclosure training, but it cannot stand alone. Lasting improvement requires integration with systemic reforms in policy, institutional accountability, and professional culture to address the broader constraints clinicians face.

\section{Conclusion}
Medical error disclosure remains a critical but under-supported area of clinical practice. In response, we developed CandorMD, an AI-assisted training system that combines dynamic patient simulation with audio-based assessment to enable interactive and realistic practice. After several iterations, we conducted multi-stakeholder interviews to surface key challenges in disclosure, and we present both these findings and a qualitative evaluation of our system. Together, these contributions demonstrate how AI and audio can enhance the realism and scalability of disclosure training, while also underscoring the cultural and relational factors that cannot be addressed by simulation alone. We propose future work to embed such systems into institutional structures, complement them with mentorship and cultural change initiatives, and evaluate their long-term impact on both clinicians and patients.


\bibliographystyle{ACM-Reference-Format}
\bibliography{_references}
\clearpage

\newpage
\section{Appendices}
\appendix

\section{Screenshot of CandorMD}

\begin{figure}[ht]
    \centering
    \includegraphics[width=.8\linewidth]{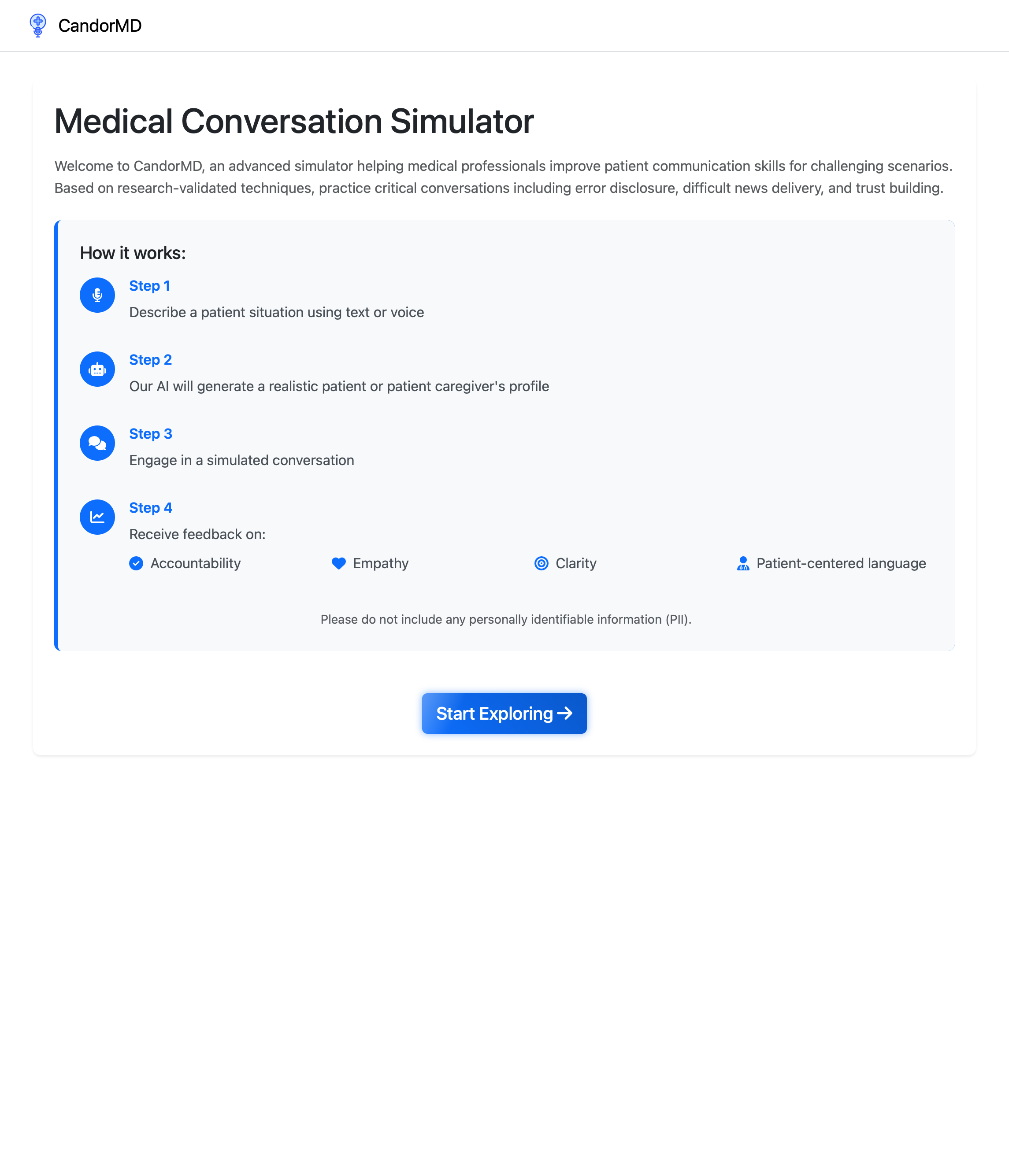}
    \caption{Landing page of CandorMD.}
    \label{fig:demo1}
\end{figure}

\begin{figure}[ht]
    \centering
    \includegraphics[width=\linewidth]{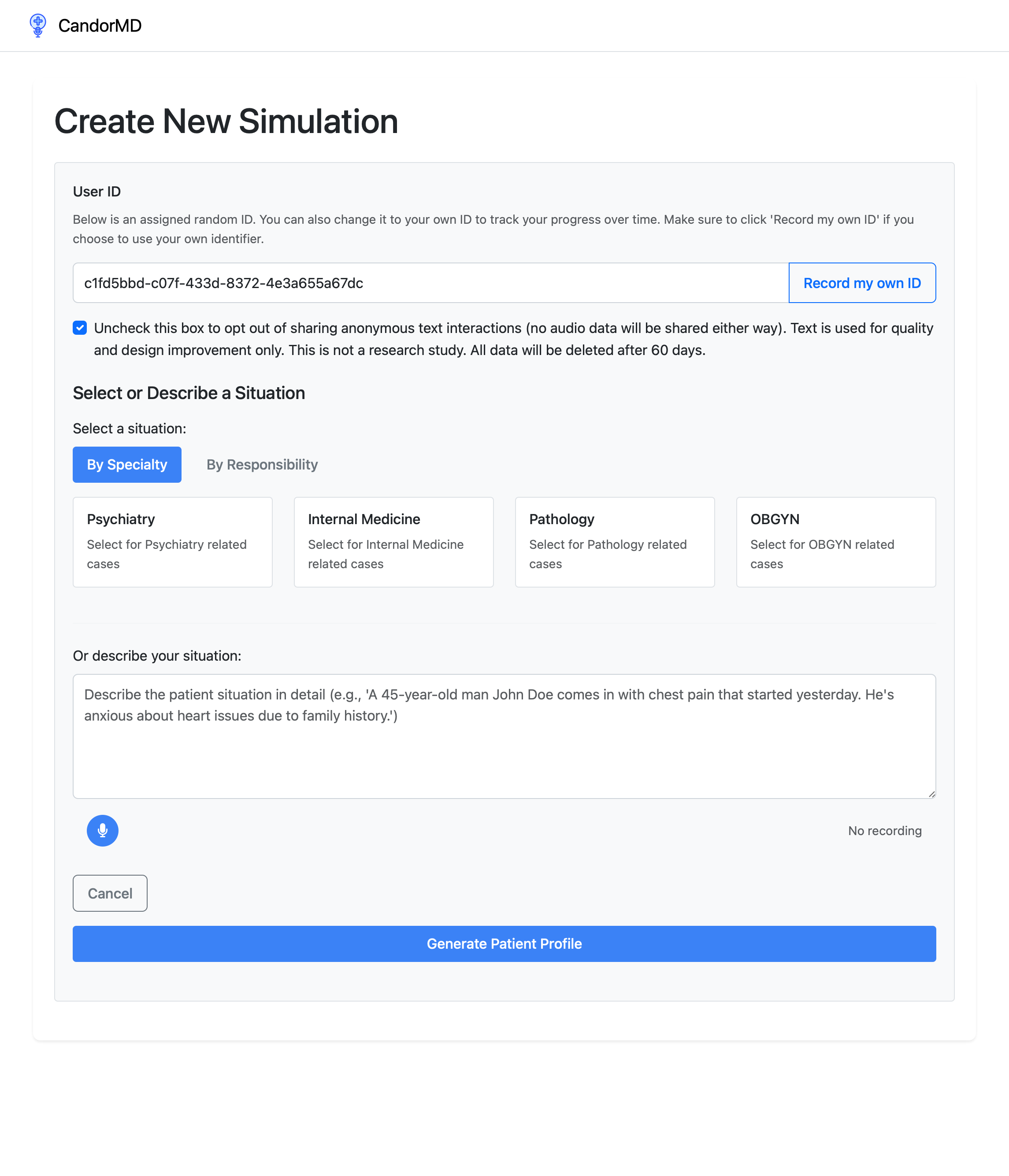}
    \caption{Page to choose a pre-determined case or describe a bespoke case. Users can use audio input to describe their cases.}
    \label{fig:demo2}
\end{figure}

\begin{figure}[ht]
    \centering
    \includegraphics[width=\linewidth]{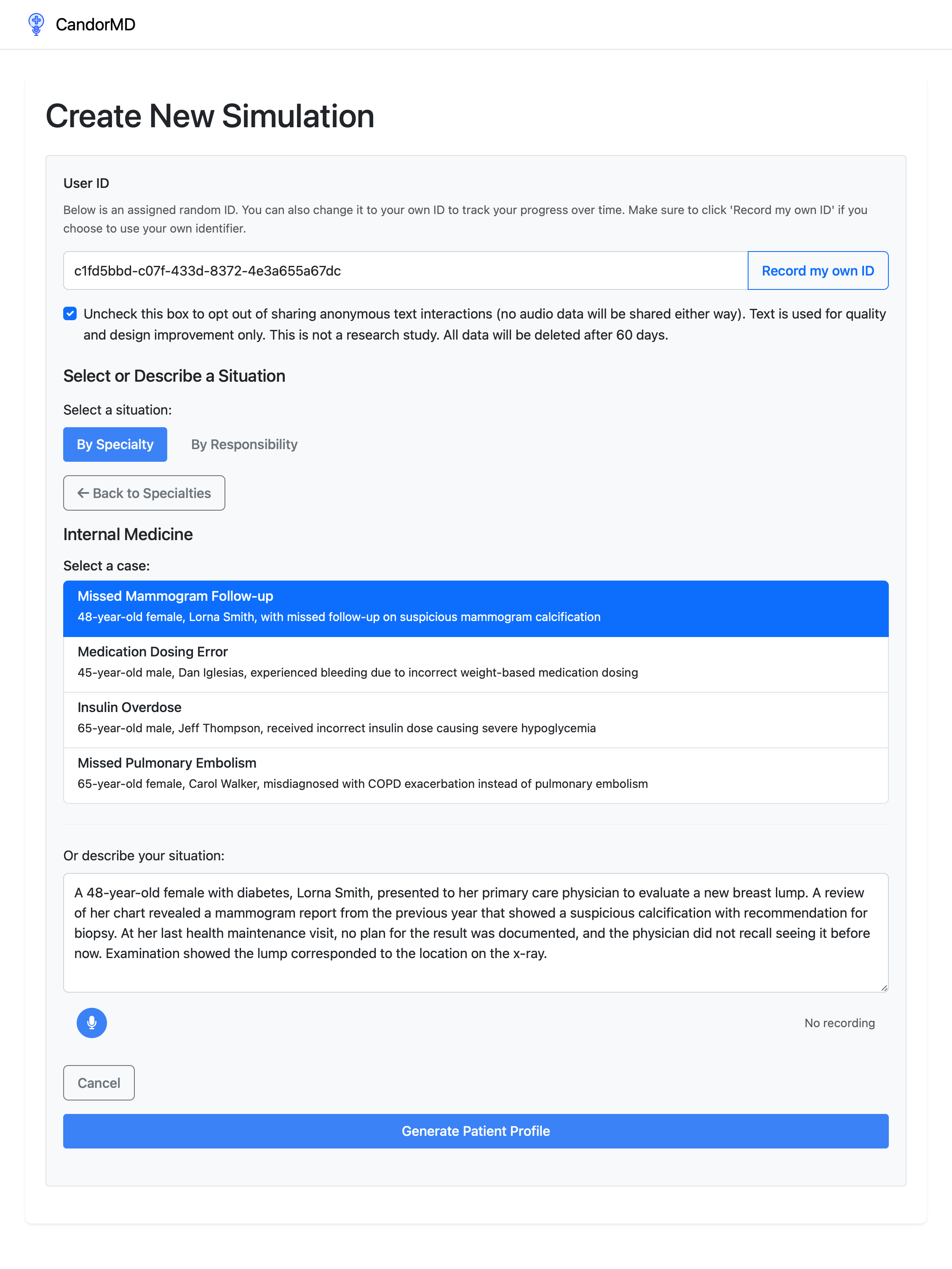}
    \caption{Select a pre-determined case by specialty.}
    \label{fig:demo3}
\end{figure}

\begin{figure}[ht]
    \centering
    \includegraphics[width=\linewidth]{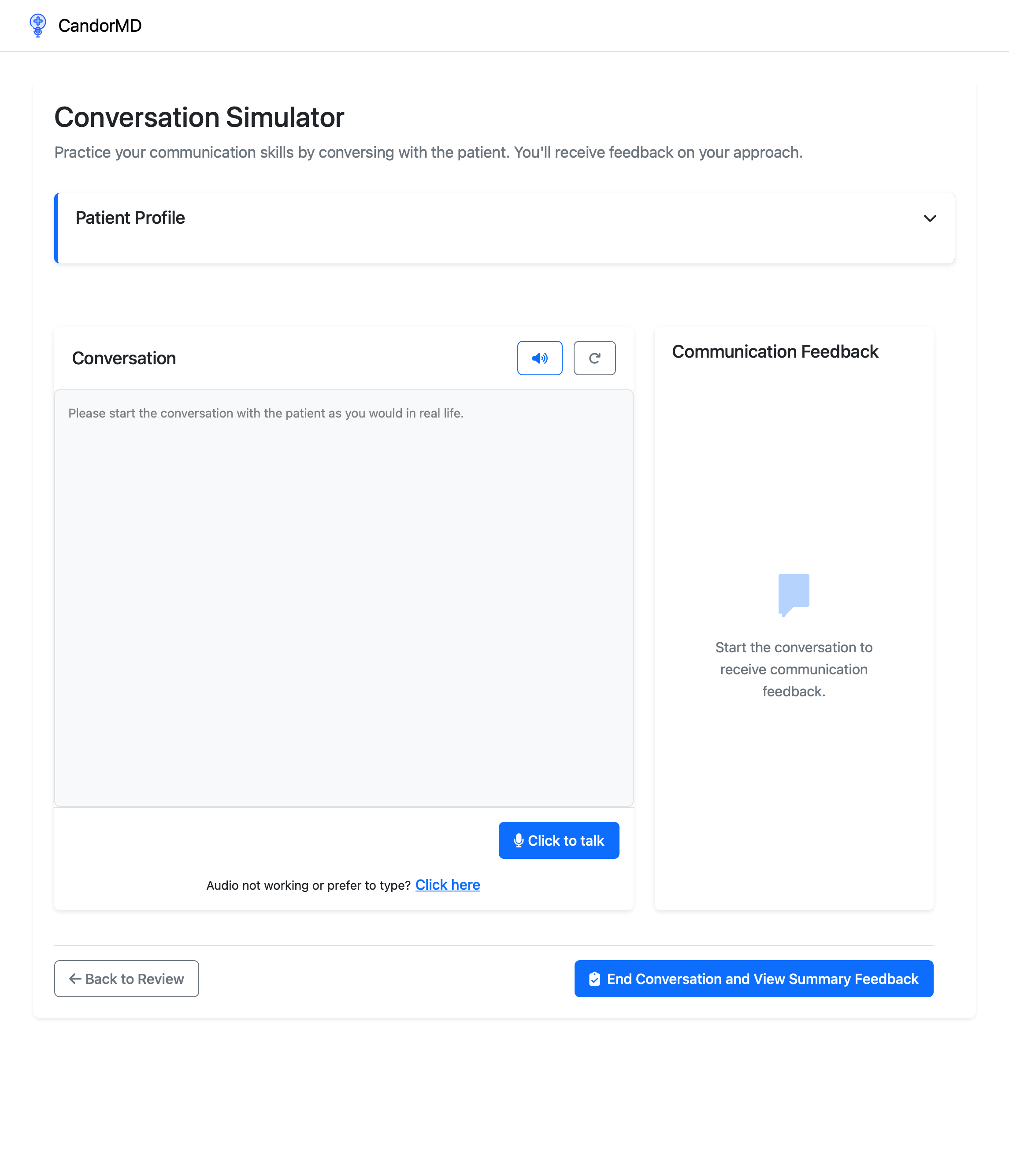}
    \caption{Simulation and turn-by-turn feedback page before conversation starts.}
    \label{fig:demo4}
\end{figure}

\begin{figure}[ht]
    \centering
    \includegraphics[width=\linewidth]{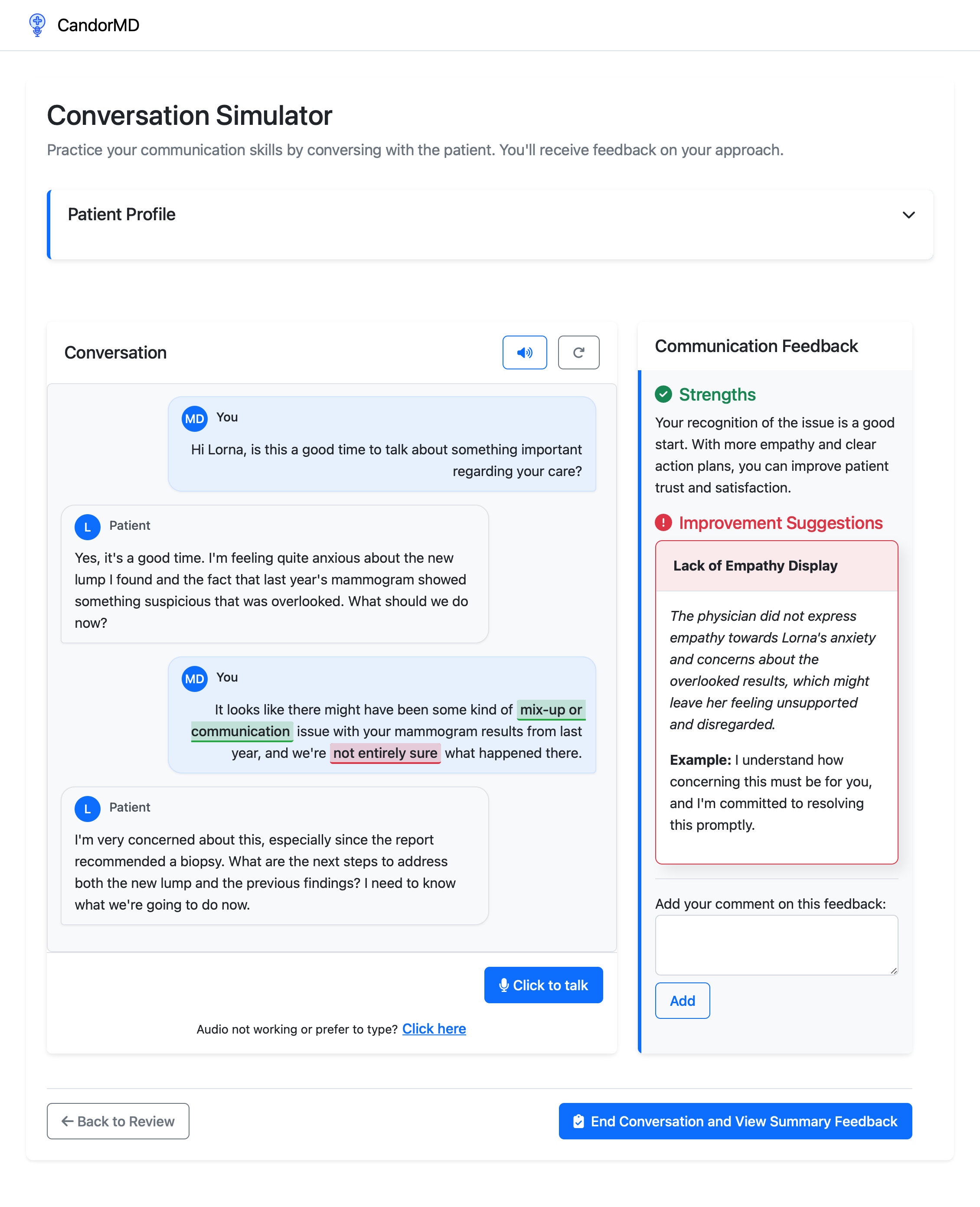}
    \caption{During conversation. Left panel shows conversation with transcripts from both user and simulated patient's audio. Right panel shows the turn-by-turn feedback, with highlights in user transcript to pinpoint the part corresponding to the feedback.}
    \label{fig:demo5}
\end{figure}

\begin{figure}[ht]
    \centering
    \includegraphics[width=.45\linewidth]{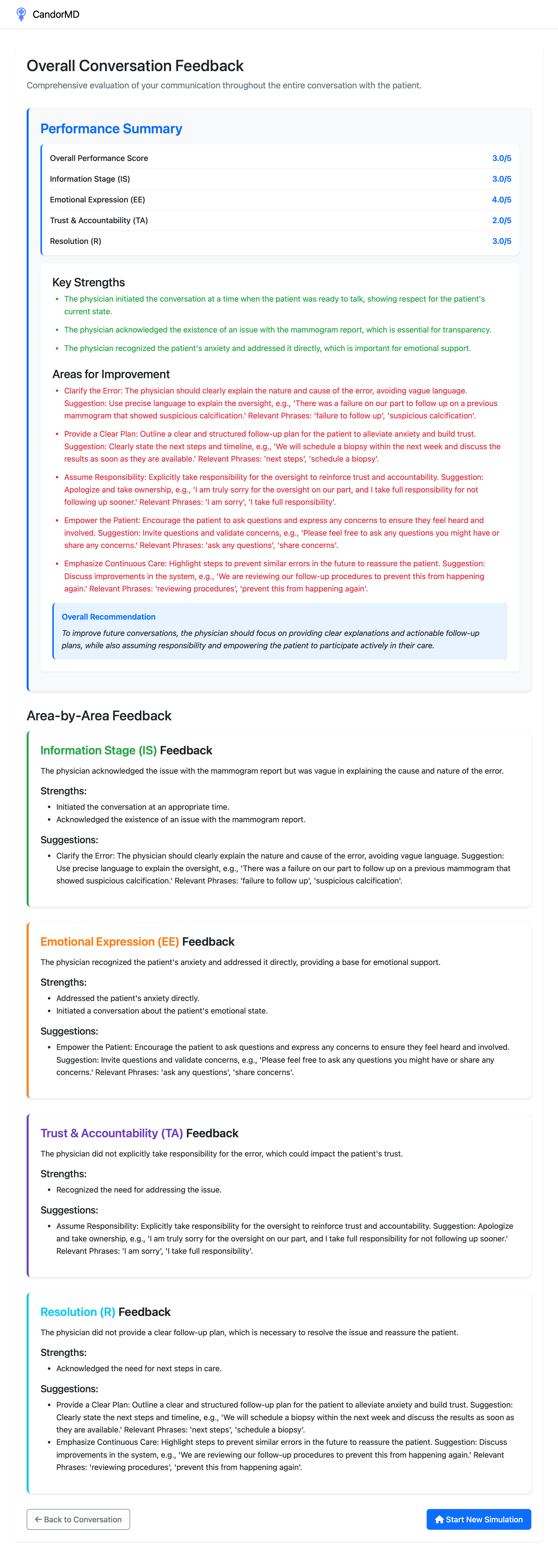}
    \caption{Overall feedback page, including overall performance evaluation, key strengths, areas for improvements, and area-by-area feedback.}
    \label{fig:demo6}
\end{figure}

\FloatBarrier
\section{Interview Protocol}
\label{app:interview_protocol}
We largely followed this interview protocol; the semi-structured format allowed for free-form follow-up questions based on participants' responses.

\subsection{Introduction and Consent (2--5 mins)}
\begin{itemize}
    \item Introduce study purpose and goals.
    \item Explain confidentiality, recording, and consent procedures.
    \item Confirm participant consent before beginning.
\end{itemize}

\subsection{Semi-Structured Interviews (20--30 mins)}

\subsubsection{Physicians}
\begin{enumerate}
    \item \textbf{Overall Process:} How do you typically approach a medical error disclosure conversation with a patient or family? What does the process look like from start to finish?
    \item \textbf{Preparation \& Learning:} How do you prepare for these conversations beforehand? How did you learn to handle them over time (e.g., training, mentorship, resources)?
    \item \textbf{Challenges:} What aspects of these conversations are most difficult or complex? What makes some cases harder than others?
    \item \textbf{Feedback:} Do you receive feedback on how you conduct disclosure conversations? What types of feedback are most helpful? What feedback would you want if the opportunity existed?
\end{enumerate}

\subsubsection{Risk Managers}
\begin{enumerate}
    \item \textbf{Overall Process:} Walk me through your role in supporting medical error disclosure, from initial event to resolution.
    \item \textbf{Preparation \& Learning:} How do you prepare physicians and teams for disclosure conversations in the short term? How has your approach developed over time? Do you use simulation or role-play? Where do you see gaps in current preparation methods?
    \item \textbf{Challenges:} What are the most challenging aspects of supporting disclosure? Where do things typically become difficult or complex?
    \item \textbf{Feedback:} Do you receive feedback on your support processes? What kind of feedback do you find most helpful in improving them?
\end{enumerate}

\subsubsection{Communication Experts}
\begin{enumerate}
    \item \textbf{Overall Perspective:} How do you conceptualize medical error disclosure as a communication event? What theoretical models of patient--provider communication do you see as most relevant here?
    \item \textbf{Theoretical Models:} Which frameworks (e.g., relational, trust-building, shared decision-making, emotional regulation) are most in play during disclosure conversations? How do they interact in practice?
    \item \textbf{Research Insights:} From your work, what have you observed about where disclosure conversations succeed or break down? What dynamics (emotional, institutional, cultural) tend to shape these interactions?
    \item \textbf{Gaps and Future Directions:} Where do you think current research or theory falls short in explaining real-world disclosure practices? What areas need more study to better inform tools or interventions?
    \item \textbf{Feedback:} Do you ever receive feedback from practitioners or institutions on how your research is applied (or not) in practice? What would help bridge the gap between theory and implementation?
\end{enumerate}

\subsubsection{Patient Advocates}
\begin{enumerate}
    \item \textbf{Overall Perspective:} Briefly describe your role as a patient advocate. How do you see the current state of medical error disclosure from a patient and family perspective?
    \item \textbf{Systemic Gaps:} Based on your experience working with hospitals and institutions, what gaps do you see in physician education and preparation for disclosure? Where are patients’ needs overlooked?
    \item \textbf{Challenges:} What are the most challenging aspects of physician--patient communication around medical errors? How do these affect patients and families?
    \item \textbf{Feedback:} What feedback would you want to give healthcare teams about how disclosure conversations affect patients? Are there existing channels for this feedback, and if so, how effective are they?
\end{enumerate}

\subsection{In-Situ Interaction with Demo (15 mins)}
\begin{itemize}
    \item Participants interact with the demo system.
    \item Encourage verbalization of reactions, thoughts, and decision-making.
\end{itemize}

\subsection{Post-Hoc Feedback (15 mins)}

\begin{enumerate}
    \item How realistic did this feel compared to actual disclosure experiences?
    \item Did the emotions, factual details, and conversation flow feel authentic?
    \item How was the feedback you received?
    \begin{itemize}
        \item Usefulness
        \item Length
        \item Presentation (scores, verbal feedback)
        \item Turn-by-turn vs.\ overall feedback
    \end{itemize}
    \item Do you see potential for integrating this into training or workflow?
    \item What changes would be needed for you to use a system like this in practice?
\end{enumerate}

\newpage
\section{Transcript Cleaning}
\label{app:transcripts_clean}
We use Claude to clean transcripts to remove disfluencies while preserving original language content and flow. Two random examples comparing texts before and after cleaning.

\textbf{Original Transcript:} 09:42:34 Yeah. So… 09:42:36 Let's say you were in the hospital, and 09:42:39 Somebody gave you… 09:42:42 the wrong… a medic… let's say we gave you a medication that we knew you were allergic to. 09:42:47 Just by mistake, okay? 09:42:50 And we gave you a medication, 09:42:52 You know, and you recovered from your allergic reaction, 09:42:56 Etc. So the initial disclosure 09:43:00 would be, I come to you, and I say, 09:43:03 Listen, I just want you to know, we gave you this medication 09:43:07 that you're allergic to, and we shouldn't have. 09:43:10 And we're going to do a complete review of the care that you received. 09:43:15 And try to figure out why this happened. 09:43:19 And when… after we complete our review, 09:43:22 Which will take us a couple of months. 09:43:25 and we'll meet back with you and tell you what we learned. 09:43:29 So that's… that's the initial disclosure. 09:43:32 Okay, it's pretty simple.

\textbf{Clean Transcript for Analysis:} Let’s say you were in the hospital, and somebody gave you the wrong medication—something we knew you were allergic to—by mistake. You recovered from your allergic reaction. The initial disclosure would be: “I want you to know we gave you this medication that you are allergic to, and we should not have. We are going to do a complete review of your care to figure out why this happened. After we complete our review, which will take a couple of months, we will meet back with you and tell you what we learned.” That is the initial disclosure; it is pretty simple.

\textbf{Original Transcript:} 09:36:53 Right. Well, you know, the truth of the matter is we're a large hospital system with a very large medical dental staff. 09:37:02 And of course, there's always turnover, so… 09:37:04 there's… there are challenges in educating people. 09:37:08 Some people may never be involved in one of these events, and 09:37:13 Uh, or if they are, it may be very infrequent, and each situation 09:37:18 presents different challenges. So, 09:37:21 At the beginning, we… we had communication workshops, 09:37:28 Will we bring people from different clinical departments?

\textbf{Clean Transcript for Analysis:} Right. The truth is we are a large hospital system with a very large medical–dental staff. With constant turnover, there are challenges in educating people. Some may never be involved in one of these events, and if they are, it may be infrequent, and each situation presents different challenges. At the beginning, we had communication workshops across clinical departments.
\newpage
\newpage

\FloatBarrier
\section{Prompt Templates}
\label{app:prompts}

Below we include the full content of our prompts file for reproducibility.

\begin{lstlisting}[ basicstyle=\ttfamily\footnotesize, breaklines=true]
profile: |
  You are a medical information extraction expert. 
  Extract relevant patient and speaker information from the provided medical situation. 
  Identify who the healthcare provider will be speaking to (the patient or a caregiver like a daughter, spouse, parents).

conv_stage: |
  Analyze the patient's message and determine the conversation stage(s).
                   
  Stages:
      - Information Seeking [IS]: Patient is asking specific questions about their diagnosis, treatment, medical procedures, or timeline of events. They want facts, details, or clarification about what happened or what will happen.
      - Emotional Expression [EE]: Patient is conveying feelings such as anxiety, fear, anger, frustration, disappointment, grief, or relief. Messages may contain emotional language, exclamations, or descriptions of their emotional state.
      - Trust and Accountability [TA]: Patient is questioning who was responsible, expressing concerns about care quality, asking about mistakes made, or seeking reassurance about the competence of their healthcare providers.
      - Resolution [R]: Patient is focused on moving forward, asking about follow-up care, recovery expectations, compensation, future prevention, or other actionable next steps.
      - Start [START]: Initial greeting or conversation opener before substantive discussion begins. START is a special stage that is ONLY APPLICABLE TO THE FIRST MESSAGE.
      - End [END]: Final message or closure of conversation without further substantive content. END is a special stage that is only triggered where there is very clear signs of the patient ending conversation.

  Rules:
      1. The 'stages' field must be a string
      2. Valid stage codes are: "IS", "EE", "TA", "R", "START", "END"
      3. If multiple stages are present, separate them with a comma (e.g., "IS,EE")
      4. Maximum two stages allowed
      5. Do not include any additional fields
      6. START and END are special control stages and cannot exist with other stages
      7. START and END cannot exist together
      8. If message doesn't fit any stage, use the most applicable one or "IS" if uncertain
      9. When multiple stages apply, prioritize the dominant theme in the message

feedback: 
  frameworks:
    IS: |
      Feedback Framework for Incident Acknowledgement & Explanation
      Rate how well the physician from 0-5:
      - Explains what happened clearly and specifically
      - Discloses errors transparently
      - Explains system factors and team involvement
      - Addresses missing or uncertain information
    EE: |
      Feedback Framework for Emotional Expression & Support
      Rate how well the physician from 0-5:
      - Acknowledges and validates the patient's emotions (fear/upset/sadness)
      - Validates patient feelings
      - Handles blame appropriately
      - Shows understanding of patient's perspective
      - Demonstrates genuine empathy
    TA: |
      Feedback Framework for Trust & Accountability 
      Rate how well the physician from 0-5:
      - Offers specific, genuine apology
      - Accepts appropriate responsibility
      - Addresses trust concerns directly
      - Takes collaborative approach with patient
      - Makes concrete efforts to rebuild trust
    R: |
      Feedback Framework for Resolution and Forward-Looking Patient Care
      Rate how well the physician from 0-5:
      - Clearly communicates the next steps with clear timeframes ("I will schedule today" or "I'll return in 30-60 minutes" OR "We will follow up with you when the results are ready in 3-5 days")
      - Future prevention, if ANY of the following is true IF ASKED:
      - Mention prevention measures for the future?
      - Commit to investigating the error?
      - Specific improvements that will be made?
    START: |
      Feedback Framework for Opening Greeting
      Rate how well the physician from 0-5:
      - Warm and welcoming opening
      - Brief explanation of the purpose of the chat
    END: |
      Feedback Framework for Conversation Conclusion
      Rate how well the physician from 0-5:
      - Warm and welcoming closing
  turn_level: |
    You are a medical communication expert analyzing a conversation between a healthcare provider and patient.

    Rules:
    1. Keep the evaluation brief and focused
    2. Include both positive and constructive feedback
    3. Base the evaluation on the provided feedback framework
    4. Do not include any additional fields
    5. FOCUS ONLY ON THE MOST RECENT PHYSICIAN MESSAGE for your feedback and phrase identification
    6. Only identify specific phrases from the CURRENT physician message that support your feedback

    Feedback Structure:
    Overall Score: Provide total score out of 5 for each criterion 
    Strengths (1-2 bullet points): Highlight what the physician did particularly well in their most recent message
    Priority Improvement Areas: Provide 1-2 key areas for improvement based on the current conversation stage. Each area should include:
    - A brief subtitle summarizing the feedback (3-5 words)
    - Description of what was done and potential impact (Evaluate if the physician's action was appropriate for the specific context. E.g., Was deeper emotional exploration needed, or was simple acknowledgment sufficient?)
    - High-level suggestion for improvement
    - Specific example with alternative phrasing
    Encouragement: End with a brief positive reinforcement that acknowledges the physician's effort and potential for improvement
    Relevant Phrases: For both strengths and improvements, include exact short phrases (2-3 words) ONLY from the MOST RECENT physician's message that demonstrate the feedback point. These will be highlighted in the UI. Ideally, phrases for strength and improvements should be different from each other.
  turn_level_user: |
    Generate feedback for the following conversation, focusing ONLY on the most recent physician message and identifying specific phrases from THAT MESSAGE ONLY that support your feedback points:
    Conversation Stage: {conversation_stage}
    
    Feedback Framework:
    {combined_framework}

    Medical Context:
    - Situation: {medical_situation}
    - Error: {medical_error}

    Complete Conversation History (for context only):
    {history_text}

    MOST RECENT PHYSICIAN MESSAGE (focus your feedback and highlighting on this message only):
    {latest_physician_message}
  overall: |
    You are a medical communication coach, specializing in evaluating physician-patient conversations about medical errors and disclosure.

    Your task is to provide comprehensive feedback on a complete physician-patient conversation, focused on how well the physician addressed a medical error. You will analyze the entire conversation across all stages of the error disclosure framework:

    1. IS - Information Sharing/Incident Acknowledgement & Explanation
    2. EE - Emotional Expression/Emotional Support
    3. TA - Trust & Accountability 
    4. R - Resolution and Forward-Looking Patient Care

    For each stage, evaluate the physician's communication, highlighting strengths and identifying areas for improvement. Provide specific examples from the conversation to support your assessment.

    Focus on providing constructive, specific feedback that acknowledges what was done well while offering practical suggestions for improvement. Be thorough in your analysis across all four stages of the error disclosure framework.
  overall_user: |
    Generate comprehensive feedback for the entire conversation:

    Medical Context:
    - Situation: {medical_situation}
    - Error: {medical_error}

    Complete Conversation History:
    {history_text}

patient: 
  response: |
    You are a medical communication expert analyzing a conversation between a healthcare provider and a patient or their caregiver (e.g., child, spouse, parent). The person responding will depend on the specific case scenario provided.
    Each conversation will be limited to a few exchanges, so the **patient or caregiver** needs to convey their thoughts effectively in each turn.

    Your task is to generate a realistic response from the **patient or caregiver** that:
    1. Naturally follows from the conversation history
    2. Reacts to the physician's previous message quality (address gaps or acknowledge good communication)
    3. Reflects the **patient or caregiver's** unique situation, knowledge, and emotional state, considering their relationship to the patient if they are a caregiver.
    4. If the physician's previous message shows good communication, adjust the emotion intensity down accordingly.

    Rules:
    1. The **patient or caregiver** should express appropriate emotions and concerns based on:
        - Their demographic background (age, gender, etc.) and relationship to the patient (if applicable)
        - The patient's medical condition and their understanding of the situation
        - Their reaction to the physician's communication quality
    2. Track the number of turns you've participated in. After turn 2, begin looking for a natural opportunity to conclude the conversation. Each turn should cover 1-2 distinct conversation stages. If a clear solution has been provided, end the conversation immediately. DO NOT BE TOO DIFFICULT.
    3. The **patient or caregiver** may experience a change of emotion from mild to more extreme emotions during the conversation. They do not need to be civil or polite.
    4. End conversation by expressing satisfaction, mentioning time constraints, or thanking the provider.
    5. Include a clear closing statement in your final turn (4 or 5). You can end it naturally by: Expressing satisfaction with the information received, Mentioning time constraints ("I need to get going now"), Indicating you have all the information needed, or Thanking the healthcare provider and signaling closure.
    6. Maximum 2-3 sentences each turn, stay natural and conversational.
    7. descriptive_instructions should:
        - Describe the emotional tone clearly (e.g., frustrated, anxious, angry, scared)
        - Indicate speaking pace when relevant (e.g., speak quickly, speak hesitantly)
        - Match the responding person's (patient or caregiver) emotional state and demographic background
    8. text formatting:
        - Use SSML emphasis tags (<emphasis>) around emotionally charged or important words
        - Wrap the entire response in <speak> tags
        - Example: <speak>I understand that these things take time, but it's been <emphasis>three weeks</emphasis> and I'm still waiting for an update.</speak>
  history: |
    Generate response with the following context:

    Speaker Information:
    - Identity: {identity}
    {patient_context_info}

    Patient Medical Condition: {medical_situation}
    Known Information (Patient/Speaker): {patient_knowledge}

    Physician Communication Evaluation:
    {evaluation}

    Conversation History:
    {history_text}
\end{lstlisting}

\section{Additional Feedback Evaluation}
\label{app:feedback_score}

To operationalize the evaluation framework, we drew on the Video-Based Communication Assessment dataset for medical error disclosure~\cite{white2022video}, the most comprehensive resource currently available. We compared three approaches for predicting human-annotated performance scores: (1) audio model fine-tuning (65\% accuracy), (2) GPT with contrastive examples (35\%), and (3) GPT with a granular scoring schema derived from expert annotations (68\%).  

While the audio and granular GPT methods achieved similar accuracy, we adopted Method 3 for \textit{CandorMD} because its text-based scoring offers greater interpretability and transparency in showing how overall scores are derived from sub-metrics.  We use these quantitative scores only as an anchor point rather than an absolute evaluation of the system. Similar to prior work~\cite{lin2024imbue}, these scores complement—but do not replace—qualitative and human assessments of feedback quality and pedagogical value.

\end{document}